\documentclass[11pt]{article}
\usepackage{amsmath,amsthm,latexsym,amssymb,amsfonts,epsfig,MnSymbol,proof}

\usepackage[all,arc,dvips,arrow,curve,cmactex]{xy}
\usepackage{amsfonts,amsmath,amssymb,graphics,psfrag}

\newcommand\ps[1]{\underline{#1}}
\newcommand\op{{\rm op}}
\newcommand\Sub{{\rm Sub}}
\newcommand\Om{\underline{\Omega}}
\newcommand\w{\mathfrak{w}}
\newcommand\ket[1]{\,|#1\rangle}
\renewcommand\l{\mathcal{L}}

\makeatletter
\@addtoreset{equation}{section}
\makeatother

\newtheorem{Definition}{Definition}[section]

\def\be{\begin{equation}}
\def\ee{\end{equation}}
\def\ba{\begin{eqnarray}}
\def\ea{\end{eqnarray}}
\newcommand\Hi{{\cal H}}
\newcommand\V{\mathcal{V}}
\newcommand\Sig{\underline{\Sigma}}
\newcommand\Sets{{\bf Sets}}

\def\mv{\mathcal{V}}
\def\mh{\mathcal{H}}

\def\us{\underline{\Sigma}}

\def\uom{\underline{\Omega}}

\def\Nl{{\mathchoice
{\setbox0=\hbox{$\displaystyle\rm N$}\hbox{\hbox to0pt
{\kern0.4\wd0\vrule height0.9\ht0\hss}\box0}}
{\setbox0=\hbox{$\textstyle\rm N$}\hbox{\hbox to0pt
{\kern0.4\wd0\vrule height0.9\ht0\hss}\box0}}
{\setbox0=\hbox{$\scriptstyle\rm N$}\hbox{\hbox to0pt
{\kern0.4\wd0\vrule height0.9\ht0\hss}\box0}}
{\setbox0=\hbox{$\scriptscriptstyle\rm N$}\hbox{\hbox to0pt
{\kern0.4\wd0\vrule height0.9\ht0\hss}\box0}}}}
\def\Zl{{\mathchoice
{\setbox0=\hbox{$\displaystyle\rm Z$}\hbox{\hbox to0pt
{\kern0.4\wd0\vrule height0.9\ht0\hss}\box0}}
{\setbox0=\hbox{$\textstyle\rm Z$}\hbox{\hbox to0pt
{\kern0.4\wd0\vrule height0.9\ht0\hss}\box0}}
{\setbox0=\hbox{$\scriptstyle\rm Z$}\hbox{\hbox to0pt
{\kern0.4\wd0\vrule height0.9\ht0\hss}\box0}}
{\setbox0=\hbox{$\scriptscriptstyle\rm Z$}\hbox{\hbox to0pt
{\kern0.4\wd0\vrule height0.9\ht0\hss}\box0}}}}
\def\Ql{{\mathchoice
{\setbox0=\hbox{$\displaystyle\rm Q$}\hbox{\hbox to0pt
{\kern0.4\wd0\vrule height0.9\ht0\hss}\box0}}
{\setbox0=\hbox{$\textstyle\rm Q$}\hbox{\hbox to0pt
{\kern0.4\wd0\vrule height0.9\ht0\hss}\box0}}
{\setbox0=\hbox{$\scriptstyle\rm Q$}\hbox{\hbox to0pt
{\kern0.4\wd0\vrule height0.9\ht0\hss}\box0}}
{\setbox0=\hbox{$\scriptscriptstyle\rm Q$}\hbox{\hbox to0pt
{\kern0.4\wd0\vrule height0.9\ht0\hss}\box0}}}}
\def\Rl{{\mathchoice
{\setbox0=\hbox{$\displaystyle\rm R$}\hbox{\hbox to0pt
{\kern0.4\wd0\vrule height0.9\ht0\hss}\box0}}
{\setbox0=\hbox{$\textstyle\rm R$}\hbox{\hbox to0pt
{\kern0.4\wd0\vrule height0.9\ht0\hss}\box0}}
{\setbox0=\hbox{$\scriptstyle\rm R$}\hbox{\hbox to0pt
{\kern0.4\wd0\vrule height0.9\ht0\hss}\box0}}
{\setbox0=\hbox{$\scriptscriptstyle\rm R$}\hbox{\hbox to0pt
{\kern0.4\wd0\vrule height0.9\ht0\hss}\box0}}}}
\def\Cl{{\mathchoice
{\setbox0=\hbox{$\displaystyle\rm C$}\hbox{\hbox to0pt
{\kern0.4\wd0\vrule height0.9\ht0\hss}\box0}}
{\setbox0=\hbox{$\textstyle\rm C$}\hbox{\hbox to0pt
{\kern0.4\wd0\vrule height0.9\ht0\hss}\box0}}
{\setbox0=\hbox{$\scriptstyle\rm C$}\hbox{\hbox to0pt
{\kern0.4\wd0\vrule height0.9\ht0\hss}\box0}}
{\setbox0=\hbox{$\scriptscriptstyle\rm C$}\hbox{\hbox to0pt
{\kern0.4\wd0\vrule height0.9\ht0\hss}\box0}}}}
\def\Hl{{\mathchoice
{\setbox0=\hbox{$\displaystyle\rm H$}\hbox{\hbox to0pt
{\kern0.4\wd0\vrule height0.9\ht0\hss}\box0}}
{\setbox0=\hbox{$\textstyle\rm H$}\hbox{\hbox to0pt
{\kern0.4\wd0\vrule height0.9\ht0\hss}\box0}}
{\setbox0=\hbox{$\scriptstyle\rm H$}\hbox{\hbox to0pt
{\kern0.4\wd0\vrule height0.9\ht0\hss}\box0}}
{\setbox0=\hbox{$\scriptscriptstyle\rm H$}\hbox{\hbox to0pt
{\kern0.4\wd0\vrule height0.9\ht0\hss}\box0}}}}
\def\Ol{{\mathchoice
{\setbox0=\hbox{$\displaystyle\rm O$}\hbox{\hbox to0pt
{\kern0.4\wd0\vrule height0.9\ht0\hss}\box0}}
{\setbox0=\hbox{$\textstyle\rm O$}\hbox{\hbox to0pt
{\kern0.4\wd0\vrule height0.9\ht0\hss}\box0}}
{\setbox0=\hbox{$\scriptstyle\rm O$}\hbox{\hbox to0pt
{\kern0.4\wd0\vrule height0.9\ht0\hss}\box0}}
{\setbox0=\hbox{$\scriptscriptstyle\rm O$}\hbox{\hbox to0pt
{\kern0.4\wd0\vrule height0.9\ht0\hss}\box0}}}}

\title{{\sf Review of the Topos Approach to Quantum Theory}\\
}

\author{{\sf C. Flori$^1$}\thanks{{\sf cflori@perimeterinstitute.ca}}\\
\\
{\sf $^1$ Perimeter Institute for Theoretical Physics,}\\
{\sf 31 Caroline Street N, Waterloo, ON N2L 2Y5, Canada}}
\date{{\small\sf June 2011 }}

\begin{document}

\maketitle
\begin{abstract}
Topos theory has been suggested by D\"oring and Isham as an alternative mathematical
structure with which to formulate physical theories. In particular, the topos approach suggests a radical new way of thinking about what a theory of physics is and what its conceptual framework looks like. The motivation of using topos theory to express quantum theory lies in the desire to overcome certain interpretational problems inherent in the standard formulation of the theory. In particular, the topos reformulation of quantum theory overcomes the instrumentalist/Copenhagen interpretation thereby rendering the theory more realist. In the process one ends up with a multivalued/intuitionistic logic rather than a Boolean logic. In this article we  shall review some of these developments.
\end{abstract}
\newpage
\section{Introduction}

One of the main challenges in theoretical physics in the past
fifty years has been to define a theory of quantum gravity, i.e., a
theory which consistently combines general relativity and quantum
theory. However, not withstanding the great effort that has been
put into this subject, physicists cannot even all
agree on what such a theory should look like. The most that has
been agreed upon is that quantum theory and general relativity should
appear as limits of the theory in the appropriate regimes.

The reasons for this elusiveness in a quantum theory of gravity
are manifold. The difficulties which arise are of two types:
`factual' and conceptual. The factual reasons are the following:
\begin{enumerate}
\item[i)] The regimes at which quantum gravity is expected to be applicable
(Planck length $10^{-35}$m and Planck energy $10^{28}$ev) are far
beyond the range of any obviously conceivable experiments. This lack of
empirical results makes it difficult to test any proposal for a
quantum theory of gravity.

\item[ii)] Given the range of potential applications of a possible quantum theory of gravity
(just after the big-bang) there is not even any agreement on what
sort of data and predictions such a theory might have.
\end{enumerate}

On the conceptual side, the problems facing quantum gravity are of
two sorts:
\begin{enumerate}
\item[i)] Conceptual obstacles that arise from the individual component theories,
i.e., general relativity and quantum theory.

\item[ii)] Conceptual obstacles that arise from trying to combine such
theories.
\end{enumerate}

The presence of such obstacles might make one wonder what actually
guides researchers in developing possible theories of quantum
gravity: i.e., how can one define a conceptual framework in a
mathematical consistent language which could represent an unknown
quantum theory of gravity for which we have no tangible
experimental evidence (beyond the limiting situations in which GR or quantum theory, respectively, apply alone). Arguably, the main guiding principle is a
philosophical prejudice of what the theory should look like,
mainly based on the success of mathematical constructs for
theories that are believed to be closely connected.

However, we will not develop such a line of thought here. Instead,
we will simply analyse a current proposal for creating a new quantum foundation for a theory
of quantum gravity. In this respect, it is interesting to note
that the different approaches to quantum gravity are based on
whether quantum theory and/or the current ideas of space and time,
i.e. GR (general relativity), are to be taken as fundamental or
not.

%
%

We will now briefly describe the conceptual problem inherent in quantum theory itself and in those arising from trying to combine GR
and quantum gravity. However, a detailed analysis of all such problems will not be given: we will just state those conceptual problems that are
most related to quantum gravity.

\begin{itemize}
\item As it stands quantum theory is non-realist. From a mathematical perspective this is reflected in the Kochen-Specker theorem\footnote{
\textbf{Kochen-Specker Theorem}: if the dimension of $\Hi$ is
greater than 2, then there does not exist any valuation function
$V:\mathcal{O}\rightarrow\Rl$ from the set
$\mathcal{O}$ of all bounded self-adjoint operators $\hat{A}$ of
$\Hi$ to the reals $\Rl$, such that  for all
$\hat{A}\in\mathcal{O}$ and all $f:\Rl\rightarrow\Rl$ the following holds $V(f(\hat{A}))=f(V(\hat{A}))$.}. This theorem implies that any statement
regarding a `state of affairs', formulated within the theory, acquires
meaning only \emph{a posteriori}, i.e. after measurement. This implies that
it is hard to avoid the Copenhagen interpretation of quantum
theory, which is intrinsically non-realist.
\item Notions of `measurement' and `external observer' pose problems when dealing with cosmology. In fact, in this case there can be no external observer since we are dealing with a closed system. But then this  implies that the concept of `measurement' plays no fundamental role, which in turn implies that the standard definition of probabilities in terms of relative frequency of measurement results breaks down.
\item The existence of the Planck scale suggests that there is no \emph{a priori} justification for the adoption of the notion of a continuum in the quantum theory used in formulating quantum gravity.
\item The formulation of standard quantum theory employs  a fixed spatio-temporal structure (fixed background). This is needed to make sense of its instrumentalist interpretation, i.e. there needs to be a space-time in which measurements can be made. This fixed background seems to cause problems in quantum gravity where one is trying to make measurements of space-time properties. In fact, if the action of making a measurement requires a space time background, what does it mean to measure space time properties?
\end{itemize}

These conceptual problems lead to the idea that maybe a new formulation of quantum theory might be needed. This is precisely what the topos approach aims at. In particular in this approach a new description of quantum theory (and hopefully in the future of quantum gravity) is put forwards, in which a radical new way of thinking about a theory of physics is proposed. At this stage
it is rather optimistic to talk about a topos proposal for a theory of quantum gravity since, to date, there is none. However, there is a
well-developed idea on how topos theory can be used in general to
describe theories of physics including, potentially, a theory of
quantum gravity.

The key idea in this approach is that constructing a theory of physics involves
finding, in a topos, a representation of a certain formal
language\footnote{A formal language is a deductive system of
reasoning made of atomic variables, relations between such
variables, and rules of inference. In this context it is assumed
that each system has a formal language attached to it and which
provides a deductive system based on intuitionistic logic (see subsequent sections). }, that
is attached to the system under investigation (see \cite{andreas1}
for a detailed analysis). Thus the topos approach consists in
first understanding at a fundamental level what a theory of
physics and associated conceptual framework should look like and,
then, applying these insights to quantum gravity. In this
context, a radically new way of thinking about space-time is
suggested: for example, the possibility that both GR and quantum
theory are `emergent' theories.

Although no topos formulation of quantum gravity has yet been
developed,
a reformulation of quantum theory and its history formulation has
recently been carried out in \cite{andreas5, andreas, andreas2,
andreas3, andreas4, 46ah} and from these works it is clear how the
Copenhagen interpretation of quantum mechanics can be replaced
with a more realist interpretation.

The details of how such a more realist interpretation is achieved
are given in subsequent sections. Here it suffices to say that the
scheme involves a synthesis of the many-worlds view and that of
extra variables. In particular, of the latter it retains the fact
that quantities have more values than those defined through the
eigenvalue-eigenstate link, while of the former it retains the
fact that these extra values are defined in terms of standard
quantum theory. This alternative interpretation of quantum theory
has been coined \emph{neo-realist}.

\section{Topos Formulation of Quantum Theory}
The conceptual problems delineated in the previous section led Isham,  D\"oring  and Butterfield to search for a
reformulation of quantum theory that is more realist\footnote{By a
`realist' theory we mean one in which the following conditions are
satisfied: (i) propositions form a Boolean algebra; and (ii)
propositions can always be assessed to be either true or false. As
will be delineated in the following, in the topos approach to
quantum theory, both of these conditions are relaxed, leading to
what Isham, Butterfield and D\"oring called  a \emph{neo-realist} theory.} than
the existing one. They achieved this through
the adoption of topos theory as the mathematical framework in
which to reformulate quantum theory.

One strategy to reformulate quantum theory in a more realist way
is to re-express it in such a way that it `looks like' classical
physics---the paradigmatic example of a realist theory. This is precisely the main idea in the topos approach: use topos theory to redefine the mathematical
structure of quantum theory in such a way that it  `looks like' classical
physics. Furthermore, this reformulation of quantum theory has the
key advantages that (i)  no fundamental role is played by the
continuum; and (ii) propositions can be given truth values without
needing to invoke the concepts of `measurement' or `observer'. This was done in
\cite{andreas1}, \cite{andreas2}, \cite{andreas3}, \cite{andreas4}
and \cite{andreas5}. A different, but related, approach to using topos theory in quantum theory was put forward in \cite{bass}. For a comparison of the two approaches see \cite{comparison}.

Thus the first question is, what is the
underlining structure that makes classical physics a realist
theory?
\\
The authors identified this structure with the following elements:
\begin{enumerate}
\item The existence of a state space $S$.

\item  Physical quantities are represented by functions from the state space to the real numbers. Thus, each physical quantity, $A$, is represented by a function $f_A:S\rightarrow \Rl$
\item Any proposition of the form ``$A\in \Delta$'' (``The value of the quantity A lies in the subset $\Delta\in\Rl$'') is represented by a subset of the state space $S$, namely that subspace for which the proposition is true. \\
This is just $f_A^{-1}(\Delta)=\{s\in S| f_A(s)\in\Delta\}$.
The collection of all such subsets forms a Boolean algebra
denoted ${\rm Sub}(S)$.

\item States $\psi$ are identified with Boolean-algebra homomorphisms $\psi:{\rm Sub}(S)\rightarrow \{0,1\}$
from the Boolean algebra ${\rm Sub}(S)$ to the two-element set
$\{0,1\}$. Here, $0$ and $1$ can be identified as `false' and
`true', respectively.
The identification of states with such maps follows from
identifying propositions with subsets of $S$. Indeed, to each
subset $f_A^{-1}(\{\Delta\})$, there is associated a
characteristic function
$\chi_{A\in\Delta}:S\rightarrow\{0,1\}\subset\Rl$ defined by
\begin{equation}
\chi_{A\in\Delta}(s)=\begin{cases}1& {\rm if}\hspace{.1in}f_A(s)\in\Delta;\\
0& \text{otherwise}  \label{eq:mam}.
\end{cases}
\end{equation}
Thus, each state $s$ either lies in $f_A^{-1}(\{\Delta\})$ or it
does not. Equivalently, given a state $s$ every proposition about
the values of a physical quantity in that state is either true or
false.
\end{enumerate}
These constructs represent the mathematical building blocks needed for a theory (in the case above, classical physics) to be considered a realist theory. However, at a deeper level, such constructs are representations of certain abstract concepts in the mathematical context of $\Sets$.
In fact, as will be explained in detail in the following sections, given a system $S$, the construction of any theory of physics regarding $S$ has, at its fundamental level, the logical structure associated to the abstract language $\l(S)$ associated to $S$. Given such a language $\l(S)$ then the definition of a theory of physics will comprise the representation of such a language in the appropriate mathematical framework.
In the case of classical physics such a mathematical context is $\Sets$.
The particular way in which the elements of the language are represented will define whether the theory is classical or not.\\

In standard quantum theory the analogues of the above mathematical constructs are:
\begin{enumerate}
\item [a.]The state space is represented by the Hilbert space $\mh$.
\item [b.] A physical quantity $A$ is represented by a self adjoint operator acting on the Hilbert space, i.e. $\hat{A}:\mh\rightarrow\mh$.
\item [c.] A proposition ``$A\in\Delta$'' is represented by the projection operator $\hat{E}[\hat{A}\in\Delta]$ that projects onto the subset $\Delta\cap sp(\hat{A})$ of the spectrum $sp(\hat{A})$ of the quantity $\hat{A}$. The collection of all such propositions forms a non-distributive logic.
\item [d.] A state $\psi$ is a vector in this Hilbert space.
\end{enumerate}
As we can see the non realism of standard quantum theory is a direct consequence of the fact that the logic of propositions is non-distributive. This, in turn, is a direct consequence of the way in which propositions are represented. Thus the aim now is to find a topos analogue for the mathematical constructs  above, which induces a distributive logic on the collection of all propositions about a system.

\subsection{Notion of a Context}

The first issue in finding quantum analogues of the constructs 1,2,3, and 4 in the previous section is to
define the appropriate mathematical framework in which to
reformulate the theory. As previously mentioned, the choice fell on
topos theory. There were many reasons for this, but a paramount
one is that in any topos (which is a special type of category)
distributive logics arise in a natural way, i.e. a topos has an
internal logical structure whose origin is similar in many ways to how Boolean algebras arise in set theory; in particular it is distributive. This feature is
highly desirable, since requirement 3 implies that the subobjects
of our state space (yet to be defined) should form some sort of distributive
logical algebra.

The second issue is to identify which  topos is the right one to
use. Isham and his collaborators achieved this by noticing that the possibility of
obtaining a `neo-realist' reformulation of quantum theory lay in
the idea of a \emph{context}. Specifically, because of the
Kochen-Specker theorem, the only way of obtaining quantum
analogues of requirements 1,2,3 and 4 is by defining them with
respect to commutative subalgebras (the `contexts') of the
non-commuting algebra, $\mathcal{B(H)}$, of all bounded operators
on the quantum theory's Hilbert space.

The set of all such commuting algebras (chosen to be von Neumann
algebras) forms a category, $\V(\Hi)$, called the \emph{context
category}. These contexts will represent classical `snapshots' of
reality, or `world-views'. From a mathematical perspective, the
reason for choosing
 commutative subalgebras as contexts is because of the existence of the Gel'fand transform which states the following:\\

\noindent
Given a commutative von Neumann
algebra V, the Gel'fand transform is a map
\begin{align}
V\rightarrow C(\Sig_V)\;;\;\;\;
\hat{A}\mapsto \bar{A}
\end{align}
which associates to each self-adjoint operator in $V$ its Gel'and transform $\bar{A}$ defined as follows:
\ba
\bar{A}:\Sig_V\rightarrow\Cl\;;\;\;\;
\lambda\mapsto\bar{A}(\lambda):=\lambda(\hat{A})
\ea
such that $im(\bar{A})=sp(\hat{A})$. Here $sp(\hat{A})$ is the spectrum of the self-adjoint operator $\hat{A}$.

In the above,  $\Sig_V$ is the \emph{Gel'fand
spectrum} of $V$, i.e. the set of all multiplicative, linear
functionals, $\lambda:V\rightarrow \Cl$, of norm 1. The existence of such a transform allows  the self-adjoint operators in a given von Neumann algebra $V$ to be written as continuous functions from the Gel'fand
spectrum to the
complex numbers. This is similar to how physical quantities are
represented in classical physics, namely as maps from the state
space to the real numbers.

In fact locally, one can interpret $\Sig_V$ as the state space and it is precisely this reason why the collection of all abelian von Neumann subalgebras has been used as the \emph{context category}. This is defined as follows:
\begin{Definition}
The category $\mv(\mh)$ of abelian von Neumann algebras has
\begin{itemize}
\item Objects: abelian von Neumann subalgebras of the algebra, $\mathcal{B(H)}$, of all bounded operators on $\Hi$.

\item Morphisms: given two algebras $V$ and $V^{'}$ there exists an arrow between them, $i_{VV^{'}}:V\rightarrow V^{'}$, iff $V\subseteq V^{'}$.
\end{itemize}
\end{Definition}
One can see that, in fact, the category $\mv(\mh)$ is a poset ordered by subset inclusion. This category is very important in the formulation of quantum theory in terms of topos theory, since it will represent contexts with respect to which any physical quantity in our formulation, is defined.

However, the meaning of contextuality that emerges in this schema is not simply that of ascribing `reality' to some
\emph{fixed} commutative subset of physical quantities, as is done in the so-called \emph{modal} approaches to quantum theory: i.e. approaches in which the value given to a physical
quantity, $A$, depends on some context in which, $A$, is to be considered
and cannot be part of a global assignment of values. Instead, in the topos approach, values of physical quantities are defined globally, but in such a way that a global definition comprises many context-dependent definitions related in a coherent way.

Now, it might look like a triviality to only consider abelian von Neumann subalgebras, but that is not the case. The reason is that the existence of arrows in the category $\mv(\mh)$ implies that it is possible to relate contexts. In particular, given two algebras $V$ and $V_1$whose intersection is non trivial, there always exist the inclusion arrows $V\cap V_1\rightarrow V$ and $V\cap V_1\rightarrow V_1$.

What this means is that a given operator $\hat{A}\in V\cap V_1$ can be written as $\hat{A}=g(\hat{B})=f(\hat{C})$, where $C\in V$, $B\in V_1$ and $f,g:\Rl\rightarrow\Rl$. Now, although it is the case that $[\hat{A}, \hat{B}]=[\hat{A},\hat{C}]=0$, it is not necessarily the case that $[\hat{B},\hat{C}]=0$. The existence of such ``Kochen-Specker triples'' $\hat{A}$, $\hat{B}$, $\hat{C}$ means that the complicated structure of quantum theory is still retained in the categorical structure of $\mv(\mh)$.
\subsection{What is Topos Theory?}
Topos theory is a vast and deep area of mathematics and it would be impossible to define it in detail in the present paper. A very good introduction to the subject is \cite{topos3}, while a more advanced exposition of the subject can be found in \cite{topos7}. In the following section we will give a very brief outline of topos theory, delineating only a few aspects of the theory that will be used later in this paper.

The very `hand-wavy' definition of a topos is a category with special properties. We will not go in the details of what these special properties are (see \cite{mythesis}): we will focus instead on what these special properties imply. Specifically, these extra properties make a topos ``look like'' the category of sets, $\Sets$, in the sense that any mathematical operation that can be done in set theory has an analogue in a general topos.

Therefore, a topos is a category for which all the categorical versions of set constructs exist and are well defined. It is precisely in this sense that a topos ``looks like'' $\Sets$. For example, there are topos analogues of the set-theoretic notions of cartesian product, $S\times T$, disjoint union, $S\coprod T$, and exponential $S^T$---the set of all functions from $T$ to $S$. Also each topos has a terminal object, denoted by $1$. This is the object with the property that, given any object, $A$, in the topos, there exists only one arrow from $A$ to $1$. In $\Sets$, the terminal object is the singleton $1 = \{*\}$.
 A \emph{(global) element} of an object $A$ is defined to be any arrow $1\rightarrow A$. This mimics the set-theory result than any point $x$ in a set $X$ can be associated with a unique function $\{*\}\rightarrow X$.

Of particular importance in a topos are the following two elements (for a detailed description the reader is refereed to \cite{topos3}, \cite{topos7}, \cite{mythesis}, \cite{andreas5}):

\textbf{Heyting algebra}.
Given a topos there exist an internal logic derived from the collection of all sub-objects of any object in the topos. This logic is  a \emph{Heyting algebra}, which is distributive but where the law of excluded middle may not hold, i.e., there exist objects $S$ such that $S\vee \neg S\leq 1$. This contrasts with the Boolean algebra in $\Sets$ for which $S\vee \neg S= 1$ for all sets $S$. An example of a Heyting algebra is given by the collection of all open sets in any topological space.

\textbf{Sub-object classifier}.
Each topos comes equipped with an object $\Omega$, called the `subobject classifier', which represents the generalisation of the set $\{0,1\}\simeq\{\text{ true, false}\}$ of truth-values in the category $\Sets$.
As the name suggests, the subobject classifier identifies subobjects. In the case of $\Sets$, given a set $A$ we say either $A\subseteq X$ or $A\nsubseteq X$. Thus to the proposition ``$A$ is a subset of $X$" we can ascribe either the value true ($1$) or false $(0)$, respectively. Moreover, if $A\subseteq X$ one can ask which points $x\in X$ lie in $A$. This can be expressed mathematically
using the characteristic function: $\chi_A:S\rightarrow\{0,1\}$, which is
defined as:
\begin{equation}\label{equ:character}
\chi_A(x)=\begin{cases}0& if\hspace{.1in}x\notin A\\
1& if\hspace{.1in}x\in A
\end{cases}
\end{equation}
where here $\{0,1\}=\{\text{false, true}\}$. Thus, in sets, we only have true or false as truth values, i.e $\Omega=\{0,1\}$. This type of truth values determines the internal logic of set to be Boolean, i.e. $S\vee\neg S=1$.

However, in a general topos $\Omega\neq\{0,1\}$ but is instead a more generalised object that leads to a multi-valued logic. In this setting we obtain a well-defined mathematical notion of what it means for an object to be `partly' a subobject of another object. Thus the role of a subobject classifier $\Omega$ in a topos is to define how subobjects fit into an object.

As was the case in $\Sets$, the (global) elements of this object $\Omega$ represent the truth values of propositions. The collection of all such truth values is a Heyting algebra. This ends our brief description of what a topos is.

There are many different kinds of topoi, and the one used in quantum theory is selected by physical motivations, in particular from the notion of context defined above. In fact we are looking for a topos that incorporates the notion of a context, such that any element in the topos can be defined locally in terms of individual abelian algebras, but such that the global information is retrieved from the categorical structure of the collection of all such algebras.
This is desirable since we are hoping for a more realist interpretation of quantum theory so as to make it `look like' classical theory. Thus, by taking into consideration abelian subalgebras, locally we have a `classical' description of the theory. However, as pointed out above, the full quantum information is retained by the categorical structure of the collection of all the abelian von Neumann subalgebras.

The topos which allows us to obtain classical local descriptions of object is the topos of presheaves over the category $\mv(\mh)$. This topos is denoted by $\Sets^{\mv(\mh)^{\op}}$.
The definition of a presheaf is as follows:

 Let $\mathcal{C},\mathcal{D}$ be categories. Then a presheaf  is an assignment to each $\mathcal{D}$-object  $A$ of a $\mathcal{C}$-object $X(A)$,
 and to each $\mathcal{D}$-arrow $f:A\rightarrow B$  a $\mathcal{C}$-arrow
$X(f):X(B)\rightarrow X(A)$, such that: i) $X(1_A)=1_{X(A)}$ and
ii) $X(f\circ g)=X(g)\circ X(f)$ for any $g:C\rightarrow
A$, i.e. a presheaf is a contravariant functor.\\
The collection of all such contravariant functors from the base category $\mv(\mh)$ to $\Sets$ forms a topos: $\Sets^{\mv(\mh)^{\op}}$.
This is the topos in which to describe quantum theory.

\section{Philosophy Behind the Topos Approach}
In this section we will try to describe the connection between topos and logic. The nature and scope of this connection is deep and wide so we will not be able to expose it in its full details. We will, however, try to give a general account of this intimate connection. The reason why such a connection is of paramount importance in our schema is that, as we will see later, it is the main ingredient in defining what a theory of physics actually is in the topos frame work.

In fact, as will be described in detail later in the paper, the definition of a theory of physics in the topos framework consists of three main steps:
\begin{enumerate}
\item [1)]
Association with each physical system $S$ of a local language, $\mathcal{L}(S)$.
\item [2)]
The application of the theory-type (for example, classical physics,
or quantum physics) to $S$. This
involves finding a representation of $\mathcal{L}(S)$ in an appropriate topos $\tau$ within whose framework the theory,
as applied to S, is to be formulated and interpreted.
\item [3)] Each topos has an internal language associated
with it, thus a theory of physics  is
equivalent to finding a translation/representation of $\mathcal{L}(S)$  into the internal language of
that topos.
\end{enumerate}
Thus the construction of a physics theory for a system $S$ is defined by an interplay between a language $\l(S)$, associated to the system $S$, a topos and the representation of the language in the topos. In particular we can say that a theory of the system $S$ is defined by choosing a representation of the language $\l(S)$ in a topos $\tau$. The choice of both topos and representation depend on the theory-type being used, i.e. if it is classical or quantum theory.
Since each topos $\tau$ has an internal language $\l(\tau)$ associated to it, constructing a theory of physics consists in translating the language, $\l(S)$, of the system in the local language $\l(\tau)$ of the topos.

\subsection{Higher-Order Typed Languages $\l$}
As explained in detail in \cite{bell} and briefly mentioned above, a topos can be seen as the representation of (higher-order) intuitionistic logic. In fact to each topos there is associated a language encoding intuitionistic logic, but also the converse is true, to each language representing intuitionistic logic there is associated a topos. Thus, we see the intimate connection between certain languages and the mathematical universe of topos theory.

So what is precisely a higher-order typed language $\l$?
A simple language, in its most raw definition, comprises a collection of atomic variables and a collection of primitive operations called logical connectives, whose role is to combine together such primitive variables transforming them into formulas or sentences. Moreover, in order to reason with a given language one also requires rules of inference, i.e. rules which allow you to generate other valid sentences from the given ones.

A higher-order typed language has some extra constructs, in particular it contains:
\begin{enumerate}
\item [i)] \emph{Types} which are particular kinds of objects, such that the primitive propositions are characterised by belonging to a certain type.
\item[ii)] $\emph{Universal quantifier }$ $\exists$ (it exists) and $\forall$ (for all).
\item[iii)] The property that quantifiers can act on individual variables, subsets of variables, functions of individual variables and various iterations of the above.
\end{enumerate}
In this setting, the formal language encodes the syntactic aspects of the theory, while the representation of such a language in a mathematical universe provides the semantics, i.e. the meaning.

In fact to actually use the language $\l$ as a deductive system of reasoning one needs to define a mathematical context in which to represent this abstract language.
In the following we will not describe in detail how the language $\l$ is defined, since it would be beyond the scope of the present paper but we will concentrate on describing the minimum set of elements present in $\l$, which makes it a suitable language for a physical system. When talking about a language as referred to a given system,  $S$, we will denote it as $\l(S)$.

In particular, given a system $S$, the minimum set of type symbols and formulas, which are needed for a language to be able to talk about $S$, are the following:
\begin{enumerate}
\item The \emph{state space object} and the \emph{quantity value object}. These are represented in $\l(S)$ by the ground type symbols $\Sigma$ and $\mathcal{R}$. Given a topos $\tau$ the object $\us$ in $\tau$ representing the state space object will be determined by physical insight and will represent the topos analogue of the state space. The representation $\underline{\mathcal{R}}$ in $\tau$ of the quantity value object may not be the real number object in a topos but something more general.

\item Given a \emph{physical quantity} $A$, we seek a representation of this object in terms of an arrow from the state space to the quantity value object. Thus, we require $\l(S)$ to contain the set function symbols $F_{\l(S)}(\Sigma, \mathcal{R})=\{\Sigma\rightarrow \mathcal{R}\}$, such that each physical quantity is represented by a symbol $A_i:\Sigma\rightarrow \mathcal{R}$. Given a topos $\tau$, these physical quantities are defined in term of $\tau$-arrows between the $\tau$-objects $\us$ and $\underline{\mathcal{R}}$, i.e. $\underline{A}:\us\rightarrow \underline{\mathcal{R}}$.

We will generally require the representation to be faithful, i.e. the map $A\rightarrow \underline{A}$ is one-to-one.

\item `Values' of physical quantities are defined in $\l(S)$ as terms of type $\mathcal{R}$ with free variables $s$ of type $\Sigma$, i.e. $A(s)$, where $A:\Sigma\rightarrow \mathcal{R}\in F_{\l(S)}(\Sigma, \mathcal{R})$.
Such terms are represented in the topos by objects\ of type $\underline{\mathcal{R}}$.
\item There is a special ground type symbol, $\Omega$, in $\l(S)$ that encodes the assignment of truth values to the internal logic. In a topos $\tau$ such an object is represented by the subobject classifier $\underline{\Omega}$ in $\tau$.\footnote{Recall that in $\Sets$ we have $\uom:=\{0,1\}$.}
\item When discussing a system we talk about its `properties': thus we deal with propositions of the form $A\in \Delta$ meaning ``the value of the quantity $A$ lies in the interval $\Delta$". Such propositions are represented in the language $\l(S)$ as terms of type $P(\Sigma)$. When representing them in a topos they are defined as elements of $P(\us)$ or, equivalently, subobjects of the state space object.

\end{enumerate}
We can now built more complicated terms of a given kind through the use of the logical connectives $\vee$, $\wedge$, $\Rightarrow$ and $\neg$.
These are called the \emph{composite formulae}.
In order for a language to be a well defined deductive system of reasoning it needs to be augmented with inference rules. Such rules enable one to deduce new formulas from existing ones. For example one has the \emph{thinning} rule which is defined as follows: given a set of formulae $\Gamma$ and individual formulae $\alpha$ and $\beta$ the thinning rule is
\[
\infer=[ ]
     {\Gamma \vdash\alpha }
     {\beta\cap\Gamma\vdash\alpha}
  \]
where the double line means double inference, i.e. the inference can be read in both directions. The symbol $\vdash$ means \emph{to imply}.

\section{Representing $\l(S)$ in the Topos $\Sets^{\mv(\mh)^{op}}$}
In this section we will show how the elements of the language $\l(S)$ are represented in the topos $\Sets^{\mv(\mh)^{op}}$ so as to describe quantum theory. The way in which these elements are represented will mimic the way in which they are represented in the classical case. The reason is that we are trying to obtain a more realist interpretation of quantum theory.

As a first element we consider the representation of the state space object $\Sig$ in $\Sets^{\mv(\mh)^{op}}$.

\subsection{Topos Analogue of the State Space}
We will now define the spectral presheaf on the category $\V(\Hi)^{op}$.
\begin{Definition}
The spectral presheaf, $\Sig$, is the covariant functor from the
category $\V(\Hi)^{op}$ to $\Sets$ (equivalently, the
contravariant functor from $\mathcal{V(H)}$ to $\Sets$) defined
by:
\begin{itemize}
\item \textbf{Objects}: Given an object $V$ in $\V(\Hi)^{op}$, the associated set $\Sig(V)=\us_V$ is defined to be the Gel'fand spectrum of the (unital) commutative von Neumann sub-algebra $V$, i.e. the set of all multiplicative linear functionals $\lambda:V\rightarrow \Cl$, such that $\lambda(\hat{1})=1$.
\item\textbf{Morphisms}: Given a morphism $i_{V^{'}V}:V^{'}\rightarrow V$ ($V^{'}\subseteq V$) in $\V(\Hi)^{op}$, the associated function $\Sig(i_{V^{'}V}):\Sig(V)\rightarrow
\Sig(V^{'})$ is defined for all $\lambda\in\Sig(V)$ to be the
restriction of the functional $\lambda:V\rightarrow\Cl$ to the
subalgebra $V^{'}\subseteq V$, i.e.
$\Sig(i_{V^{'}V})(\lambda):=\lambda_{|V^{'}}$.
\end{itemize}
\end{Definition}

\subsection{Example}
We will now describe a simple example of the topos analogue of the state space, other examples can be found in \cite{dasain}. Let us consider a four-dimensional Hilbert space $\mh=\Cl^4$. The first step is to identify the poset of abelian von-Neumann subalgebras $\mv(\Cl^4)$. Such algebras are subalgebras of the algebra $\mathcal{B}(\Cl^4)$ of all bounded operators on $\mh$. Since the Hilbert space is $\Cl^4$, $\mathcal{B}(\Cl^4)$ is the algebra of all $4\times 4$ matrices with complex entries which act as linear transformations on $\Cl^4$.
In order to form the abelian von -Neumann subalgebras one considers on orthonormal basis $(\psi_1,\psi_2,\psi_3,\psi_4)$ and projection operators $(\hat{P}_1, \hat{P}_2, \hat{P}_3, \hat{P}_4)$, which project on the one-dimensional subspaces $\Cl\psi_1$, $\Cl\psi_2$, $\Cl\psi_3$, $\Cl\psi_4$, respectively. A von-Neumann subalgebra $V$ is, then, generated by the double commutant\footnote{Consider a Hilbert space $\mh$ whose algebra of bounded operators is $B(\mh)$. Given a subalgebra $N\subset B(\mh)$, the commutant $N^{'}$ is
\be
N^{'}:=\{\hat{A}\in B(\mh)|[\hat{A},\hat{B}]=0\;\forall\; \hat{B}\in N\}\subset B(\mh)
\ee
The double commutant of  $N$ is then $(N^{'})^{'}=N^{''}$.} of collections of the above projection operators, i.e. $V=lin_{\Cl}(\hat{P}_1, \hat{P}_2, \hat{P}_3, \hat{P}_4)$.

In matrix notation, the representatives for the projection operators are
\[\hat{P}_1=\begin{pmatrix} 1& 0& 0&0\\
0&0&0 &0\\
0&0&0&0\\
0&0&0&0
  \end{pmatrix}\;\;\;\;
\hat{P}_2=\begin{pmatrix} 0& 0& 0&0\\
0&1&0 &0\\
0&0&0&0\\
0&0&0&0
  \end{pmatrix}\;\;\;\;
\hat{P}_3=\begin{pmatrix} 0& 0& 0&0\\
0&0&0 &0\\
0&0&1&0\\
0&0&0&0
  \end{pmatrix}\;\;\;\;
  \hat{P}_4=\begin{pmatrix} 0& 0& 0&0\\
0&0&0 &0\\
0&0&0&0\\
0&0&0&1
  \end{pmatrix}
\]
The largest abelian von-Neumann subalgebra generated by the above projectors is $V=lin_{\Cl}(\hat{P}_1, \hat{P}_2, \hat{P}_3, \hat{P}_4)$, i.e. the algebra consisting of all $4\times 4$ diagonal matrices with complex entries on the diagonal. Since this algebra is the largest, i.e. not contained in any other abelian subalgebra of $\mathcal{B}(\Cl)$, it is called \emph{maximal}.

Any change of basis $(\psi_1,\psi_2,\psi_3,\psi_4)\rightarrow (\rho_1, \rho_2, \rho_3, \rho_4)$ would give another maximal von-Neumann subalgebra $V_{1}=lin_{\Cl}(\rho_1, \rho_2, \rho_3, \rho_4)$. In fact there are uncountably many such maximal algebras. If two basis are related by a simple permutation or phase factor, then the abelian von-Neumann subalgebras they generate are the same.\\
Now, considering again our example, the algebra $V$ will have many non maximal subalgebras which, however, can be divided into two kinds as follows:
\be
V_{\hat{P}_i\hat{P}_j}:=lin_{\Cl}(\hat{P}_i, \hat{P}_j, \hat{P}_k+\hat{P}_l)=\Cl\hat{P}_i+\Cl \hat{P}_j+\Cl (\hat{P}_k+\hat{P}_l)=\Cl\hat{P}_i+\Cl \hat{P}_j+\Cl(\hat{1}- \hat{P}_i+\hat{P}_j)
\ee
for $i\neq j\neq k \neq l\in\{1,2,3,4\}$\\
and
\be
V_{\hat{P}_i}:=lin_{\Cl}(\hat{P}_i, \hat{P}_j+\hat{P}_k+\hat{P}_l)=\Cl\hat{P}_i+\Cl(\hat{P}_j+\hat{P}_k+\hat{P}_l)=\Cl\hat{P}_i+\Cl(\hat{1}-\hat{P}_i)
\ee
for $i\neq j\neq k \neq l=1,2,3,4$\\\\
Again, there are uncountably many non-maximal abelian subalgebras. It is the case, though, that different maximal subalgebras have common non-maximal subalgebras, and ditto for non-maximal abelian subalgebras. Thus, for example, the context $V$ above contains all the subalgebras $V_{\hat{P}_i\hat{P}_j}$ and $V_{\hat{P}_i}$ for $i,j\in\{1,2,3,4\}$. Now, consider another four, pairwise orthogonal, projection operators $\hat{P}_1$,  $\hat{P}_2$,  $\hat{Q}_3$,  $\hat{Q}_4$, such that the maximal abelian von Neumann algebra $V_{1}:=lin_{\Cl}(\hat{P}_1,\hat{P}_2, \hat{Q}_3,\hat{Q}_4)\neq V$. We then have that
$
V\cap V_{1}=\{V_{\hat{P}_1}, V_{\hat{P}_2}, V_{\hat{P}_1,\hat{P}_2}\}
$.

From the above discussion it is easy to deduce that the subalgebras $V_{\hat{P}_i}$ are contained in all the other subalgebras which contain the projection operator $\hat{P}_i$, and the subalgebras $V_{\hat{P}_i\hat{P}_j}$ are contained in all the subalgebras which contain both projection operators $\hat{P}_i$ and $\hat{P}_j$.

We should also mention that there is the trivial algebra $V^{0}:=\Cl\hat{1}$, but we will not include this algebra in $\mv(\mh)$ for reasons that will appear later later on.

Now that we have defined the category $\mv(\Cl^2)$ we will define the spectral presheaf.
Let us first consider the maximal abelian subalgebra $V=lin_{\Cl}(\hat{P}_1, \hat{P}_2, \hat{P}_3, \hat{P}_4)$. Then the Gel'fand spectrum $\us_V$ (which has a discrete topology) of this algebra contains four elements $\{\delta_1,\delta_2,\delta_3,\delta_4\}$ where
\be
\lambda_i(\hat{P}_j)=\delta_{ij}\;\;\;\; (i=1,2,3,4)
\ee
We then consider the subalgebra $V_{\hat{P}_1\hat{P}_2}=lin_{\Cl}(\hat{P}_1,\hat{P}_2, \hat{P}_3+\hat{P}_4)$. Its Gel'fand spectrum $\us_{V_{\hat{P}_1\hat{P}_2}}$ contains the elements
\be
\us_{V_{\hat{P}_1\hat{P}_2}}=\{\lambda^{'}_1, \lambda^{'}_2, \lambda^{'}_3 \}
\ee
such that $\lambda^{'}_1(\hat{P}_1)=1$, $\lambda^{'}_2(\hat{P}_2)=1$ and $\lambda^{'}_3(\hat{P}_3+\hat{P}_4)=1$, with the rest being zero.

Since $V_{\hat{P}_1\hat{P}_2}\subseteq V$, there exists a morphisms between the respective spectra as follows (for notational simplicity we will denote $V_{\hat{P}_1\hat{P}_2}$ as $V^{'}$)
\be
\us_{VV^{'}}:\us_V\rightarrow \us_{V^{'}}\;;\;\;\;\;\;
\lambda\mapsto\lambda_{|V^{'}}
\ee
such that:
\be
\us_{VV^{'}}(\lambda_1)=\lambda_1^{'}\;;\;\;\;
\us_{VV^{'}}(\lambda_2)=\lambda_2^{'}\;;\;\;\;
\us_{VV^{'}}(\lambda_3)=\lambda_3^{'}\;;\;\;\;
\us_{VV^{'}}(\lambda_4)=\lambda_3^{'}
\ee

From the above simple example we can generalise the definition of the spectrum for all subalgebras $V^{'}\subseteq V$. In particular we get
\be
\us_{V_{\hat{P}_i,\hat{P}_j}}=\{\lambda_i,\lambda_j,\lambda_{kl}\}\text{ where }\lambda_i(\hat{P}_j)=\delta_{ij}\;,\;\;\
\lambda_{kl}(\hat{P}_k+\hat{P}_l)=1
\ee
and  the rest are zero. On the other hand for contexts $V_{\hat{P}_i}$ we obtain:
\be
\us_{V_{\hat{P}_i}}=\{\lambda_i,\lambda_{jkl}\}\text{ where }\lambda_i(\hat{P}_i)=1\;\;,\;\; \lambda_{jkl}(\hat{P}_j+\hat{P}_k+\hat{P}_k)=1
\ee
and the rest are zero.

\section{Propositions}
We will now describe how certain terms of type $P(\Sigma)$ are represented in $\Sets^{\mv(\mh)^{op}}$, namely propositions. These are represented by projection operators in quantum theory and in the topos theory they are identified with clopen subobjects of the spectral presheaf.
 A \emph{clopen} subobject
$\ps{S}\subseteq\Sig$ is an object such that, for each context
$V\in \mathcal{V(H)}^{\op}$, the set $\ps{S}(V)$ is a clopen (both
closed and open) subset of $\Sig(V)$, where the latter is equipped
with the usual compact and Hausdorff spectral topology.
We will now show, explicitly, how propositions are defined.

As a first step we introduce the concept of
`daseinisation' \cite{andreas1}. Roughly speaking, daseinisation
approximates operators so as to `fit' into any given context $V$.
In fact, because the formalism defined so far is
contextual, any  proposition  one wants to consider has to be
studied within (with respect to) each context $V\in\V(\Hi)$.

To see how this works let us consider the projection operator $\hat{P}$, which corresponds, via the
spectral theorem, to the proposition ``$A\in\Delta$''\footnote{It should be noted that different propositions may correspond to the same projection operator, i.e. the mapping from propositions to projection operators is many to one. Thus, to account for this, to each projection operator one is really associating an equivalence class of propositions. The reason why von Neumann algebras were chosen instead of general C$^*$ algebras is precisely because all projections representing propositions are contained in the former, but not necessarily in the latter.}. In
particular, let us take a context $V$ such that $\hat{P}\notin
P(V)$ (the lattice of projection operators in $V$). We somehow need to define a
projection operator which does belong to $V$ and which is related,
in some way, to our original projection operator $\hat{P}$. This
can be achieved (\cite{andreas1}, \cite{andreas2}, \cite{andreas3},
\cite{andreas4} and \cite{andreas5}) by approximating $\hat{P}$
from above in $V$, with the `smallest' projection operator in $V$,
greater than or equal to $\hat{P}$. More precisely, the
\emph{outer daseinisation},
 $\delta^o(\hat P)$, of $\hat P$ is defined at each context $V$ by
 \begin{equation}
\delta^o(\hat{P})_V:=\bigwedge\{\hat{R}\in
P(V)|\hat{R}\geq\hat{P}\}
\end{equation}

This process of outer daseinisation takes place for all contexts
and hence gives, for each projection operator $\hat{P}$, a
collection of daseinised projection operators, one for each
context V, i.e.,
\begin{align}
\hat{P}\mapsto\{\delta^o(\hat{P})_V|V\in\V(\Hi)\}
\end{align}
Because of the Gel'fand transform, to each operator $\hat{P}\in
P(V)$ there is associated the map $\bar{P}:\Sig_V\rightarrow\Cl$,
which takes values in $\{0,1\}\subset\Rl\subset\Cl$ since
$\hat{P}$ is a projection operator. Thus, $\bar{P}$ is a
characteristic function of the subset
$S_{\hat{P}}\subseteq\Sig(V)$ defined by
\begin{equation}
S_{\hat{P}}:=\{\lambda\in\Sig(V)|\bar{P}(\lambda):=\lambda(\hat{P})=1\}
\end{equation}
Since $\bar{P}$ is continuous with respect to the spectral
topology on $\underline\Sigma(V)$, then
$\bar{P}^{-1}(1)=S_{\hat{P}}$ is a clopen subset of
$\underline\Sigma(V)$, since both $\{0\}$ and $\{1\}$ are clopen
subsets of the Hausdorff space $\Cl$.

Through the Gel'fand transform it is then possible to define a
bijective map between projection operators, $\delta(\hat{P})_V\in
P(V)$, and clopen subsets of the spectral presheaf $\Sig_V$. Namely,
for each \emph{context} $V$,
\begin{equation}
S_{\delta^o(\hat{P})_V}:=\{\lambda\in\Sig_V|\lambda
(\delta^o(\hat{P})_V)=1\}
\end{equation}

This correspondence between projection operators and clopen
subsets of the spectral presheaf $\underline\Sigma$ implies the
existence of a lattice homeomorphism for each $V$
\begin{equation}\label{equ:smap}
\mathfrak{S}:P(V)\rightarrow \Sub_{cl}(\Sig)_V\text{ such that }\delta^o(\hat{P})_V\mapsto
\mathfrak{S}(\delta^o(\hat{P})_V):=S_{\delta^o(\hat{P})_V}
\end{equation}

It was shown in \cite{andreas1}, \cite{andreas2}, \cite{andreas3},
\cite{andreas4} and \cite{andreas5} that the collection of subsets
$\{S_{\delta(\hat{P})_V}\mid V\in\mathcal{V(H)}\}$ is a subobject
of $\Sig$. This enables us to define the (outer) daseinisation as
a mapping from the projection operators to the subobject of the
spectral presheaf given by
\begin{align}\label{ali:glob}
\delta:P(\Hi)\rightarrow \Sub_{cl}(\Sig)\;;\;\;\;
\hat{P}\mapsto(\mathfrak{S}(\delta^o(\hat{P})_V))_{V\in\V(\Hi)}=:\ps{\delta(\hat{P})}
\end{align}
We will sometimes denote $\mathfrak{S}(\delta^o(\hat{P})_V)$ as
$\ps{\delta(\hat{P})}_V$.

Since the subobjects of the spectral presheaf form a Heyting
algebra, the above map associates propositions to a distributive
lattice. Actually, it is first necessary to show that the
collection of \emph{clopen} subobjects of $\underline\Sigma$ is a
Heyting algebra, but this was done in \cite{andreas5}.


\subsection{Understanding Daseinisation}
What does it exactly mean to daseinise a projection? Let us consider a projection $\hat{P}$ which represents the proposition $A\in\Delta$. We now consider a context $V$ such that $\hat{P}\notin V$, thus we approximate this projection so as to be in $V$ obtaining $\delta^o(\hat{P})_V$. If the projection $\delta^o(\hat{P})_V$ is a spectral projector of the operator, $\hat{A}$, representing the quantity, $A$, than it represents the proposition $A\in\Gamma$ where $\Delta\subseteq \Gamma$.

If , on the other hand, $\delta^o(\hat{P})_V$ is not a spectral projector of the operator $\hat{A}$ than $\delta^o(\hat{P})_V$ represents some proposition $B\in\Delta^{'}$.  Given the fact that $\hat{P}\leq\delta^o(\hat{P})_V$, the proposition $B\in\Delta^{'}$ is a coarse graining of $A\in\Delta$, in fact a general form of $B\in\Delta^{'}$ could be $f(A)\in\Gamma$, for some Borel function $f: sp(\hat{A})\rightarrow \Rl$.
Therefore, the mapping
\ba
\delta^o_V:P(\mh)\rightarrow P(\mv)\;;\;\;\;
\hat{P}\mapsto\delta^o(\hat{P})_V
\ea
is the mathematical implementation of the idea of \emph{coarse-graining} of propositions. Obviously, for many contexts it is the case that $\delta^o(\hat{P})_V=\hat{1}$.

From the analysis above we can deduce that, in this framework, there are two types of propositions:
\begin{enumerate}
\item[i)]\emph{Global propositions}, which are the propositions we start with and which we want to represent in  various contexts, i.e. $(A\in\Delta)=\hat{P}$.
\item[ii)]\emph{Local propositions}, which are the individual coarse graining of the global propositions, as referred to individual contexts $V$, i.e. $\underline{\delta(\hat{P})}_V$
\end{enumerate}
Thus, for every global proposition we obtain a collection of local propositions
\be
\hat{P}\rightarrow (\delta^o(\hat{P})_V)_{V\in\V(\Hi)}
\ee
In the topos perspective we consider the collection of all these local propositions at the same time, as exemplified by equation \ref{ali:glob}.

\subsection{Example}
To illustrate the concept of daseinisation of propositions let us consider a spin-2 system (a three dimensional similar example can be found in \cite{dasain}). We are interested in the spin in the $z$ direction which is represented by the physical quantity $S_z$. In particular, we want to consider the following proposition $S_z\in [1.3,2.3]$. Since the total spin in the $z$ direction can only have values $-2$, $0$, $2$, the only value in the closed interval $ [1.3,2.3]$ which $S_z$ can take is $2$.\\
The self-adjoint operator representing $S_z$ is thus
\[\hat{S}_z=\begin{pmatrix} 2& 0& 0&0\\
0&0&0 &0\\
0&0&0&0\\
0&0&0&-2
  \end{pmatrix}
  \]
 The eigenstate with eigenvalue 2 is $\psi=(1,0,0,0)$, whose associated projector $\hat{P}:=\hat{E}[S_z\in [1.3,2.3]]=|\psi\rangle\langle\psi|$ is
  \[\hat{P}_1=\begin{pmatrix} 1& 0& 0&0\\
0&0&0 &0\\
0&0&0&0\\
0&0&0&0
  \end{pmatrix}
  \]
From our definition of $\mv(\Cl^4)$ we know that the operator $\hat{S}_z$ is contained in all algebras which contain the projector operators $\hat{P}_1$ and $\hat{P}_4$. These algebras are: i) the maximal algebra, denoted $V$; and ii) the non-maximal subalgebra $V_{\hat{P}_1\hat{P}_4}$. We will now analyse how the proposition $S_z\in [1.3,2.3]$, represented by the projection operator $\hat{P}_1$, gets represented in the various abelian von Neumann algebras in $\mv(\Cl^4)$.
\begin{enumerate}
\item \emph{Context $V$ and its subalgebras}.

Since $V$, $V_{\hat{P}_1\hat{P}_i}$ ( $i\in\{2,3,4\}$) and $V_{\hat{P}_1}$ contain the projection operator $P_1$, then, for all these contexts we have
\be
\delta^o(\hat{P}_1)_V=\delta^o(\hat{P}_1)_{V_{\hat{P}_1\hat{P}_i}}=\delta^o(\hat{P}_1)_{V_{\hat{P}_1}}=\hat{P}_1
\ee
On the other hand, for context $V_{\hat{P}_i}$ for $i\neq 1$ we have
\be
\delta^o(\hat{P}_1)_{V_{\hat{P}_i}}=\hat{P}_1+\hat{P}_j+\hat{P}_k\;\;\; j\neq i\neq k\in\{2,3,4\}
\ee
For contexts of the form $V_{\hat{P}_i\hat{P}_j}$, where $i\neq j\neq1$, we have
\be
\delta^o(\hat{P}_1)_{V_{\hat{P}_i\hat{P}_j}}=\hat{P}_1+\hat{P}_k\;\;\; j\neq i\neq k\in\{2,3,4\}
\ee
\item\emph{Other maximal algebras which contain $\hat{P}_1$ and their subalgebras}.\\
Let us consider four pairwise-orthogonal projection operators $\hat{P_1},\hat{Q}_2,\hat{Q}_3,\hat{Q}_4$, such that the maximal abelian von Neumann algebra generated by such projections is different from $V$, i.e.
$$V_1:=lin_{\Cl}( \hat{P_1},\hat{Q}_2,\hat{Q}_3,\hat{Q}_4)\neq V$$
We then have the following daseinised propositions.
For contexts $V_1$ and $V_{\hat{P}_1}$, as before, we have
\be
\delta^o(\hat{P}_1)_{V_1}=\delta^o(\hat{P}_1)_{V_{\hat{P}_1}}=\hat{P}_1
\ee
for contexts $V_{\hat{Q}_i}$ we have
\be
\delta^o(\hat{P}_1)_{V_{\hat{Q}_i}}=\hat{P}_1+\hat{Q}_j+\hat{Q}_k\;\;i\neq j\neq k\in\{2,3,4\}
\ee
while, for context $V_{\hat{Q}_i\hat{Q}_j}$, we have
\be
\delta^o(\hat{P}_1)_{V_{\hat{Q}_i\hat{Q}_j}}=\hat{P}_1+\hat{Q}_k\;\;i\neq j\neq k\in\{2,3,4\}
\ee
\item \emph{Contexts which contain a projection operator which is implied by $\hat{P}_1$}.

Let us consider contexts $\tilde{V}$ which contain a projection operator $\hat{Q}$, such that $\hat{Q}\geq \hat{P}_1$, but do not contain $\hat{P}_1$ (if they did contain $\hat{P}_1$, we would be in exactly the same situation as above). In this situation the daseinised propositions will be
\be
\delta^o(\hat{P}_1)_{\tilde{V}}=\hat{Q}
\ee
\item\emph{Context which contains neither $\hat{P}_1$ or a projection operator implied by it}.\\
In these contexts $V_2$ the only coarse-grained proposition related to $\hat{P}_1$ is the unity operator. Therefore we have
\be
\delta^o(\hat{P}_1)_{V_2}=\hat{1}
\ee
\end{enumerate}
Now that we have defined all the possible coarse-grainings of the proposition $\hat{P}_1$, for all possible contexts, we can define the presheaf $\ps{\delta(\hat{P}_1)}$ which is the topos analogue of the proposition $S_z\in [1.3,2.3]$. To this end we must apply the daseinisation map\footnote{Note that so far we have only used the outer daseinisation.} defined in \ref{ali:glob}, obtaining
\ba
\delta:P(\Cl)\rightarrow \Sub_{cl}(\Sig)\;;\;\;
\hat{P}_1\mapsto(\mathfrak{S}(\delta^o(\hat{P}_1)_V))_{V\in\V(\Cl^4)}=:\ps{\delta(\hat{P})}
\ea
where the map $\mathfrak{S}$ was defined in \ref{equ:smap}, in particular
\be
\mathfrak{S}(\delta^o(\hat{P}_1)_V):=S_{\delta^o(\hat{P}_1)_V}:=\{\lambda\in\Sig_V|\lambda
(\delta^o(\hat{P}_1)_V)=1\}
\ee
We now want to define the $\ps{\delta(\hat{P}_1)}$-morphisms. In order to do so we will subdivide our analysis in different cases as above.
\begin{enumerate}
\item \emph{Maximal algebra $V$ and its subalgebras}.\\
The subalgebras of $V$ are of two kinds: $V_{\hat{P}_i,\hat{P}_j}$ and $V_{\hat{P}_k}$ for $i,j,k\in\{1,2,3,4\}$, thus the $\ps{\delta(\hat{P}_1)}$-morphisms with domain $\ps{\delta(\hat{P}_1)}_V$ will be of two kinds. We will analyse one at the time.
\be
\ps{\delta(\hat{P}_1)}_{V,V_{\hat{P}_i,\hat{P}_j}}:\ps{\delta(\hat{P}_1)}_{V}\rightarrow \ps{\delta(\hat{P}_1)}_{V_{\hat{P}_i,\hat{P}_j}}
\ee
In this context we have
\be
\ps{\delta(\hat{P}_1)}_{V}=\{\lambda_1|\lambda_1(\delta(\hat{P}_1)_{V})=\lambda_1(\hat{P})=1\}=\lambda_1
\ee
This is the case since, as we saw in the previous section $\us_V=\{\lambda_1, \lambda_2\lambda_3\lambda_4\}$ where $\lambda_i(\hat{P}_j)=\delta_{ij}$.\\
On the other hand for the contexts $V_{\hat{P}_i,\hat{P}_j}$ $i,j\in\{1,2,3,4\}$ we have the following:
\ba
\ps{\delta(\hat{P}_1)}_{V_{\hat{P}_1,\hat{P}_j}}&=&\{\lambda_1\} \text{ where }\lambda_1\Big(\delta(\hat{P}_1)_{V_{\hat{P}_1,\hat{P}_j}}=\hat{P}\Big)=1;\;\;j\in\{2,3,4\}\\
\ps{\delta(\hat{P}_1)}_{V_{\hat{P}_i,\hat{P}_j}}&=&\{\lambda_{1k}\} \text{ where }\lambda_{1k}\Big(\delta(\hat{P}_1)_{V_{\hat{P}_i,\hat{P}_j}}=(\hat{P}_1+\hat{P}_k)\Big)=1;\;\; i\neq j \neq k\in \{2,3,4\}
\ea
The $\ps{\delta(\hat{P}_1)}$-morphisms for the above contexts would be
\be
\ps{\delta(\hat{P}_1)}_{V,V_{\hat{P}_1,\hat{P}_j}}(\lambda_1):=\lambda_1\;;\;\;\;
\ps{\delta(\hat{P}_1)}_{V,V_{\hat{P}_i,\hat{P}_j}}(\lambda_1):=\lambda_{1k}
\ee
The remaining $\ps{\delta(\hat{P}_1)}$-morphisms with domain $\ps{\delta(\hat{P}_1)}_V$ are
\be
\ps{\delta(\hat{P}_1)}_{V,V_{\hat{P}_i}}:\ps{\delta(\hat{P}_1)}_{V}\rightarrow \ps{\delta(\hat{P}_1)}_{V_{\hat{P}_i}}
\ee
In this case the local representative $\ps{\delta(\hat{P}_1)}_{V_{\hat{P}_i}}$, $i\in\{1,2,3,4\}$ are
\be
\ps{\delta(\hat{P}_1)}_{V_{\hat{P}_1}}=\{\lambda_1\}\;;\;\;\;
\ps{\delta(\hat{P}_1)}_{V_{\hat{P}_i}}=\{\lambda_{1jk}\}\;\;i,j,k\in\{2,3,4\}
\ee
The $\ps{\delta(\hat{P}_1)}$-morphisms are then
\be
\ps{\delta(\hat{P}_1)}_{V,V_{\hat{P}_1}}(\lambda_1):=\lambda_1\;;\;\;\;
\ps{\delta(\hat{P}_1)}_{V,V_{\hat{P}_i}}(\lambda_1):=\lambda_{1kl}
\ee
\item\emph{Other maximal algebras which contain $\hat{P}_1$ and their subalgebras}.

As before we consider four pairwise-orthogonal projection operators $\hat{P_1},\hat{Q}_2,\hat{Q}_3,\hat{Q}_4$, such that the maximal abelian von Neumann algebra generated by them is different from $V$, i.e.

$lin_{\Cl}( \hat{P_1},\hat{Q}_2,\hat{Q}_3,\hat{Q}_4)\neq V$.
We then obtain the following morphisms with domain $\ps{\delta(\hat{P}_1)}_{V}$:
\be
\ps{\delta(\hat{P}_1)}_{V,V_{\hat{P}_1}}:\ps{\delta(\hat{P}_1)}_{V}\rightarrow \ps{\delta(\hat{P}_1)}_{V_{\hat{P}_1}}\;;\;\;\;
\lambda_1\mapsto\lambda_1
\ee
\be
\ps{\delta(\hat{P}_1)}_{V,V_{\hat{Q}_i}}:\ps{\delta(\hat{P}_1)}_{V}\rightarrow \ps{\delta(\hat{P}_1)}_{V_{\hat{Q}_i}}\;;\;\;\;
\lambda_1\mapsto\rho_{1jk}
\ee
where $\us_{\hat{Q}_i}:=\{\rho_i,\rho_{1jk}$, such that $\rho_i(\hat{Q}_i)=1$ and $\rho_{1jk}(\hat{P}_1+\hat{Q}_j+\hat{Q}_k)=1$.
\ba
\ps{\delta(\hat{P}_1)}_{V,V_{\hat{Q}_i,\hat{Q}_j}}:\ps{\delta(\hat{P}_1)}_{V}\rightarrow \ps{\delta(\hat{P}_1)}_{V_{\hat{Q}_i,\hat{Q}_j}}\;;\;\;\;
\lambda_1\mapsto\rho_{1k}
\ea
where $\us_{\hat{Q}_i,\hat{Q}_j}:=\{\rho_i,\rho_j,\rho_{1k}\}$, such that $\rho_i(\hat{Q}_i)=1$, $\rho_j(\hat{Q}_j)=1$ and $\rho_{1k}(\hat{P}_1+\hat{Q}_k)=1$.
The computation of the remaining maps is straightforward.
\item\emph{Contexts which contain a projection operator which is implied by $\hat{P}_1$}.

We now consider a context $\tilde{V}$ which contains an operator $\hat{Q}$, such that $\hat{Q}\geq \hat{P}$.

For such a context we have $\ps{\delta(\hat{P}_1)}_{\tilde{V}}=\{\lambda|\lambda(\hat{Q})=1\}$. Therefore, for subalgebras which contain the operator $\hat{Q}$ the morphisms will simply map $\lambda$ to itself. The rest of the maps are easily derivable.
\item\emph{Context which neither contain $\hat{P}_1$ or a projection operator implied by it}.

In such a context $V_2$ , whatever its spectrum is, each of the multiplicative linear functionals $\lambda_i\in\us_{V_2}$ will assign value 1 to $\ps{\delta(\hat{P}_1)}=\hat{1}$. And so will the elements of the spectrum of the subalgebras of $V_2$. Thus, all the maps $\ps{\delta(\hat{P}_1)}_{V_2, \bar{V}}$ will simply be equivalent to spectral presheaf maps.
\end{enumerate}

\section{Representation of the Type Symbol $\Omega$}
In the topos $\Sets^{\mv(\mh)^{op}}$ the representation of the linguistic object $\Omega$ is identified with the following presheaf.
\begin{Definition}
The presheaf $\Om\in \Sets^{\V(\Hi)^{op}}$ is defined as follows:
\begin{enumerate}
\item For any $V\in\mathcal{V(H)}$, the set $\Om(V)$ is defined as the set of all sieves (see definition \ref{def:sieve}) on $V$.

\item Given a morphism $i_{V^{'}V}:V^{'}\rightarrow V$ $(V^{'}\subseteq V)$, the associated function in $\underline\Omega$ is $\Om(i_{V^{'}V}):\Om(V)\rightarrow \Om(V^{'})$; $S \mapsto \Om((i_{V^{'}V}))(S):=\{V^{''}\subseteq V^{'}|V^{''}\in S\}$.
\end{enumerate}
\end{Definition}\label{def:sieve}
The definition of a sieve on a poset, in our case $\V(\Hi)$, is as follows:
\begin{Definition}
For all $V\in\V(\Hi)$, a sieve $S$ on $V$ is a collection of
subalgebras $V^{'}\subseteq V$ such that, if $V^{'}\in S$ and
$V^{''}\subseteq V^{'}$, then $V^{''}\in S$. Thus $S$ is a
downward closed set.
\end{Definition}
\noindent
In this case a maximal (principal) sieve on $V$ is $\downarrow\! V:=\{V^{'}\in\V(\Hi)|V^{'}\subseteq V\}.
$

As previously stated, truth values are identified with global
elements of the presheaf $\Om$, i.e. a collection for each $V$ of local elements in $\uom_V$ that match up in an appropriate way. The global element that consists
entirely of principal sieves is interpreted as representing
`totally true'. In classical Boolean logic this is just  `true'.
Similarly, the global element that consists of empty sieves  is
interpreted as `totally false'. In classical Boolean logic this
is just `false'.

A very important property of sieves is that the set $\uom_V$  of sieves on $V$ has the
structure of a Heyting algebra where the unit element $\underline{1}_{\uom_V}\in \uom_V$ is represented by the principal sieve $\downarrow\! V$ and, the null element $\underline{0}_{\uom_V}\in \uom_V$, is represented by the empty set $\emptyset$.
Moreover $\uom_V$ is equipped with a partial ordering given by subset inclusion, such that $S_i\leq S_j$ iff $S_i\subseteq S_j$.

An important result is that the collection of (global) elements of $\uom$ has a Heyting algebra structure.

\subsection{Example}
We will now describe an example of the truth object for the case of our four-dimensional Hilbert space $\mh=\Cl^4$. Let us start with the \emph{maximal algebras} $V=lin_{\Cl}(\hat{P}_1,\hat{P}_2, \hat{P}_3, \hat{P}_4)$. What follows can be generalised to any maximal subalgebra, not just $V$.

The collection of sieves on $V$ is
\be
\uom_V:=\{S, S_{12}, S_{13}, S_{13}, S_{23}, S_{24}, S_{34}, S_{1},S_{2}, S_{3}, S_{4}, \underline{0}_{\uom_V}\}
\ee
where each $S_{ij}$ is defined as follows:
\be
S=\{V, V_{\hat{P}_i,\hat{P}_j}, V_{\hat{P}_k}, | i,j,k\in\{1,2,3,4,\} \}\;;\;\;
S_{ij}=\{V_{\hat{P}_i,\hat{P}_j}, V_{\hat{P}_i}, V_{\hat{P}_j}\}\;;\;\;
S_{i}=\{V_{\hat{P}_i}\}\;;\;\;
\underline{0}_{\uom_V}=\{\emptyset\}
\ee
We now consider a non maximal-algebra $V_{\hat{P}_i,\hat{P}_j}$. The collection of sieves on such an algebra is\\
$
\uom_{V_{\hat{P}_i,\hat{P}_j}}:=\{S_{ij}, S_i, S_j, \underline{0}_{V_{\hat{P}_i,\hat{P}_j}}\}
$,
where the definitions of the individual sieves are the same as before.
Similarly, for the context $V_{\hat{P}_i}$ we have
\be
\uom_{V_{\hat{P}_i}}:=\{S_{i}\}
\ee
We now want to define the $\uom$-morphisms. To this end, let us first consider the $\uom$-morphism with domain $\uom_V$. There are various such morphisms $\uom_{V,V_{\hat{P}_i,\hat{P}_j}}:\uom_V\rightarrow \uom_{V_{\hat{P}_i,\hat{P}_j}}$, one for each pair $i,j\in\{1,2,3,4\}$, which is defined component-wise as follows:
\ba
S\mapsto S_{ij}\;;\;\;
S_{ij}\mapsto S_{ij}\;;\;\;
S_{ik}\mapsto S_i\;;\;\;
S_{kj}\mapsto S_j\;;\;\;
S_i \mapsto S_i\;;\;\;
S_{kl}\mapsto\underline{0}_{V_{\hat{P}_i,\hat{P}_j}}\;;\;\;
S_j\mapsto S_j\;;\;\;
S_k \mapsto \underline{0}_{V_{\hat{P}_i,\hat{P}_j}}
\ea
Moreover, for each $k\in \{1,2,3,4\}$ we have the $\uom$-morphisms $
\uom_{V,V_{\hat{P}_k}}:\uom_V\rightarrow \uom_{V_{\hat{P}_k}}$ which component-wise are defined as follows:
\ba
S\mapsto S_{k}\;;\;\;
S_{ij}\mapsto \underline{0}_{V_{\hat{P}_k}}\;;\;\;
S_{ik}\mapsto S_k\;;\;\;
S_{l}\mapsto\underline{0}_{V_{\hat{P}_k}}\;;\;\;
S_k \mapsto S_k\;;\;\;
S_j\mapsto \underline{0}_{V_{\hat{P}_k}}\;;\;\;
S_i\mapsto \underline{0}_{V_{\hat{P}_k}}
\ea
It is straightforward to extend the definition of $\uom$-morphisms for all contexts $V_i\in \mv(\Cl^4)$.

\section{States}
 In classical physics a pure state, $s$, is a point in the state
space.  It is the smallest subset of the state space which has
measure one with respect to the Dirac measure $\delta_s$. This is
a consequence of the one-to-one correspondence, which subsists
between pure states and Dirac measure. In particular, for each
pure state, $s$, there corresponds a unique Dirac measure
$\delta_s$. Moreover, propositions which are true in a pure state
$s$ are given by subsets of the state space which have measure one,
with respect to the Dirac $\delta_s$, i.e. those subsets which
contain, $s$,. The smallest such subset is the one-element set
$\{s\}$. Thus, a pure state can be identified with a single point
in the state space.
More general states are represented by more
general probability measures on the state space. This is the
mathematical framework that underpins classical statistical
physics.

However, the spectral presheaf $\Sig$ has \emph{no}
points\footnote{Recall that in a topos $\tau$, a `point' (or `global element'
or just `element') of an object $O$ is defined to be a morphism
from the terminal object, $1_\tau$, to $O$.}. Indeed, this is
equivalent to the Kochen-Specker theorem! Thus the analogue of a
pure state must  be identified with some other construction. There
are two (ultimately equivalent)\ possibilities:  a `state' can be
identified with (i) an element of $P(P(\Sig))$; or (ii) an element
of $P(\Sig)$. The first choice is called the \emph{truth-object}
option, the second is  the \emph{pseudo-state} option. In what
follows we will concentrate on the second option only. For an analysis of the first option see \cite{andreas5}.

Given a pure quantum state $\psi\in\Hi$, we define
the presheaf
\begin{equation}
\ps{\w}^{\ket\psi}:= \ps{\delta(\ket\psi\langle\psi|)}
\end{equation}
so that for each stage V we have
\begin{equation}
\ps{\delta(\ket\psi\langle\psi|)}_V:=
\mathfrak{S}(\bigwedge\{\hat{\alpha}\in
P(V)|\ket\psi\langle\psi|\leq\hat{\alpha}\}) \subseteq\Sig(V)
\end{equation}
where the map $\mathfrak{S}$ was defined in equation
(\ref{equ:smap}).

It was shown in \cite{andreas1}, \cite{andreas2}, \cite{andreas3},
\cite{andreas4} and \cite{andreas5} that the map
\begin{equation}
\ket\psi\rightarrow \ps{\w}^{\ket\psi}
\end{equation}
is injective. Thus, for each state $\ket\psi$, there is associated a
topos pseudo-state, $\ps{\w}^{\ket\psi}$, which is defined as a
subobject of the spectral presheaf $\Sig$.

This presheaf $\ps{\w}^{\ket\psi}$ is interpreted as the smallest
clopen subobject of $\Sig$, which represents the proposition which
is totally true in the state $\psi$. Roughly speaking, it is the
closest one can get to defining a point in $\Sig$.
\subsection{Example}
We will now give an example of pseudo-states in our four-dimensional Hilbert space $\Cl^4$. This is very similar to the example for propositions, since also for pseudo-states the concept of daseinisation is utilised. However, for pedagogical reasons we will, nonetheless report it below.

Let us consider a state $\psi=(0,1,0,0)$. The respective projection operator is
\[\hat{P}_2=|\psi\rangle\langle\psi|=\begin{pmatrix} 0& 0& 0&0\\
0&1&0 &0\\
0&0&0&0\\
0&0&0&0
  \end{pmatrix}
  \]
We now want to compute the outer daseinisation of such a projection operator for various contexts $V\in \mv(\Cl)$. As was done for propositions, we will subdivide our analysis into different cases:
\begin{enumerate}
\item \emph{Context $V$ and its subalgebras}.

Since the maximal algebra $V$ is such that $|\psi\rangle\langle\psi|\in P(V)$, it follows that:
\be
\delta^o(|\psi\rangle\langle\psi|)_V=|\psi\rangle\langle\psi|
\ee
This also holds for any subalgebra of $V$ containing $|\psi\rangle\langle\psi|$, i.e., $V_{\hat{P}_2, \hat{P}_i}$ for $i=\{1,3,4\}$ and $V_{\hat{P}_2}$.
On the other hand, for the algebras $V_{\hat{P}_i,\hat{P}_j}$, where $i, j=\{1,3,4\}$, we have
\be
\delta^o(|\psi\rangle\langle\psi|)_{V_{\hat{P}_i,\hat{P}_j}}=\hat{P}_2+\hat{P}_k\;\; \text{for } k\neq i \neq j
\ee
For contexts $V_{\hat{P}_i}$ for $i\in\{1,3,4\}$ we have
\be
\delta^o(|\psi\rangle\langle\psi|)_{V_{\hat{P}_i}}=\hat{P}_2+\hat{P}_k+\hat{P}_j\;\; \text{for } k\neq i \neq j
\ee
\item\emph{Other maximal algebras which contain $|\psi\rangle\langle\psi|$ and their subalgebras}.

Consider the four pairwise-orthogonal projection operators $(\hat{Q}_1,\hat{P}_2, \hat{Q}_3,\hat{Q}_4)$, such that $V_1:=lin_{\Cl}( \hat{Q}_1,\hat{P}_2, \hat{Q}_3,\hat{Q}_4)\neq V$. For these contexts we obtain
\ba
\delta^o(|\psi\rangle\langle\psi|)_{V_1}=\hat{P}_2\;;\;\;
\delta^o(|\psi\rangle\langle\psi|)_{V_{\hat{Q_i}}}=\hat{P}_2+\hat{Q}_j+\hat{Q}_k\;;\;\;
\delta^o(|\psi\rangle\langle\psi|)_{V_{\hat{Q_i}.\hat{Q}_j}}=\hat{P}_2+\hat{Q}_k
\ea
\item\emph{Contexts which contain a projection operator which is implied by $|\psi\rangle\langle\psi|$}.

If $V_2$ contains $\hat{Q}\geq|\psi\rangle\langle\psi|$ then
$
\delta^o(|\psi\rangle\langle\psi|)_{V_2}=\hat{Q}
$.
\item\emph{Contexts which contain neither $|\psi\rangle\langle\psi|$ nor a projection operator implied by it}. We gave

\be
\delta^o(|\psi\rangle\langle\psi|)_{V^{'}}=\hat{1}
\ee
\end{enumerate}
The $\ps{\w}^{\ket\psi}$-morphisms  are defined in an analogous way to the $\ps{\delta(\hat{P})}$-morphisms.

\section{Topos Representation of the Type Symbol $\mathcal{R}$}
In the topos $\Sets^{\mv(\mh)}$ the representation of the quantity value object $\mathcal{R}$ is given by the following presheaf:
\begin{Definition}
The presheaf $\ps{\Rl^{\leftrightarrow}}$ has as
\begin{enumerate}
\item [i)] Objects\footnote{A map $\mu:\downarrow\! V\rightarrow\Rl$ is said to be \emph{order preserving} if $V^{'}\subseteq V$ implies that $\mu(V^{'})\leq\mu(V)$. A map $\nu:\downarrow\! V\rightarrow\Rl$ is \emph{order reversing} if $V^{'}\subseteq V$ implies that $\nu(V^{'})\supseteq \nu(V)$.}:
$
\ps{\Rl^{\leftrightarrow}}_V:=\{(\mu,\nu)\mid\mu,\nu:\downarrow V\rightarrow\Rl; \text{ where } \mu\text{ is order preserving, }\nu\text{ is order reversing and } \mu\leq\nu\}
$.
\item[ii)] Arrows: given two contexts $V^{'}\subseteq V$ the corresponding morphism is

$
\ps{\Rl^{\leftrightarrow}}_{V,V^{'}}:\ps{\Rl^{\leftrightarrow}}_V\rightarrow\ps{\Rl^{\leftrightarrow}}_{V^{'}}$;
$(\mu,\nu)\mapsto(\mu_{|V^{'}},\nu_{|V^{'}})$.

\end{enumerate}
\end{Definition}
The presheaf $\ps{\Rl^{\leftrightarrow}}$ is the object in which  physical quantities take their `values'. Thus it plays the same role as the real numbers in classical physics.

The reason why the quantity value object is defined in terms of order-reversing and order-preserving functions is because, in general, in quantum theory one can only assign approximate values to physical quantities. In particular, in most cases, the best approximation to the value of a physical quantity one can give is the smallest interval of possible values of that quantity.

Let us analyse the presheaf $\ps{\Rl^{\leftrightarrow}}$ in more depth. To this end we assume that we want to define the value of a physical quantity $A$, given a state $\psi$. If $\psi$ is an eigenstate of $A$, then we would get a sharp value of the quantity $A$, say $a$. If $\psi$ is not an eigenstate, then we would get a certain range $\Delta$ of values for $A$, where $\Delta\subseteq sp(\hat{A})$.

Let us assume that $\Delta=[a,b]$. Then what the presheaf $\ps{\Rl^{\leftrightarrow}}$ does is to single out the end points $a$ and $b$, so as to give a range (unsharp) of values for the physical quantity $A$. Obviously, since we are in the topos of presheaves, we have to define each object contextually, i.e., for each context $V\in\mv(\mh)$.
It is precisely to accommodate this fact that the pair of order reversing and order preserving functions was chosen to define the extreme values of our intervals.

To understand this consider a context $V$ such that the self-adjoint operator $\hat{A}$, which represents the physical quantity $A$, belongs to $V$ and such that the range of values of $A$ at $V$ is $[a,b]$.
If we then consider the context $V^{'}\subseteq V$, such that $\hat{A}\notin V$, we will have to approximate $\hat{A}$ so as to fit $V^{'}$. The precise way in which self-adjoint  operators are approximated will be described later, however, such an approximation will inevitably coarse-grain $\hat{A}$, i.e., it will make it more general.

It follows that the range of possible values of such an approximated operator $\hat{A}_1$ will be bigger.
Therefore the range of values of $\hat{A}_1$ at $V^{'}$ will be, $[c,d]\supseteq [a,b]$, where $c\leq a$ and $d\geq b$.
These relations between the extremal points can be achieved by the presheaf $\ps{\Rl^{\leftrightarrow}}$ through the order-reversing and order-preserving functions.
Specifically, given that $a:=\mu(V)$, $b:=\nu(V)$ since $V^{'}\subseteq V$, it follows that $c:=\mu(V^{'})\leq\mu(V)$ ($\mu$ being order preserving) and $d:=\nu(V^{'})\geq \nu(V)$ ($\nu$ being order reversing). Moreover, the fact that, by definition, $\mu(V)\leq \nu(V)$, it implies that as one goes to smaller and smaller contexts the intervals $(\mu(V)_i,\nu(V)_i)$ keep getting bigger or stay the same. An example of the quantity value object can be found in \cite{dasain}.

\section{Topos Representation of Physical Quantities}
We will now define the representation of the type functions $A:\Sigma\rightarrow \mathcal{R}$ (which in the language $\l(S)$ represent  physical quantities) as functors, in the topos $\Sets^{\mv(\mh)^{op}}$. Since the quantity value object in $\Sets^{\mv(\mh)^{op}}$ is represented by the presheaf $\ps{\Rl^{\leftrightarrow}}$, it follows that $A:\Sigma\rightarrow \mathcal{R}$ is represented by
$
\breve{\delta}(\hat{A}):\us\rightarrow \ps{\Rl^{\leftrightarrow}}
$
which, at each context $V$, is defined as
\ba
\breve{\delta}(\hat{A})_V:\us_V\rightarrow \ps{\Rl^{\leftrightarrow}}_V\;;\;\;
\lambda\mapsto(\mu_{\lambda},\nu_{\lambda})
\ea
In order to understand the precise way in which the functor $\breve{\delta}(\hat{A})$ is defined, we need to introduce the notion of \emph{spectral order}. The reason why this order was chosen rather than the standard operator ordering\footnote{Recall that the standard operator ordering is given as follows: $\hat{A}\leq\hat{B}$ iff $\langle\psi|\hat{A}|\psi\rangle\leq\langle\psi|\hat{B}|\psi\rangle$ for all $|\psi\rangle$.} is because the former preserves the relation between the spectra of the operator, i.e., if $\hat{A}\leq_s\hat{B}$, then $sp(\hat{A})\subseteq sp(\hat{B})$. This feature will be very important when defining the values for physical quantities.

We will now define what the spectral order is. Consider two self-adjoint operators $\hat{A}$ and $\hat{B}$ with spectral families $(\hat{E}^{\hat{A}}_r)_{r\in\Rl}$ and $(\hat{E}^{\hat{B}}_r)_{r\in\Rl}$, respectively. Then the spectral order is defined as follows:
\be
\hat{A}\leq_s\hat{B}\;\;\text{ iff }\;\;\forall r\in\Rl\;\;\;\hat{E}^{\hat{A}}_r\geq\hat{E}^{\hat{B}}_r
\ee
From the definition it follows that the spectral order implies the usual order between operators, i.e. if  $\hat{A}\leq_s\hat{B}$ then $\hat{A}\leq\hat{B}$, but the converse is not true.

We are now ready to define the  functor $\breve{\delta}(\hat{A})$. To this end, let us consider the self-adjoint operator $\hat{A}$ and a context $V$, such that $\hat{A}\notin V_{sa}$ ($V_{sa}$ denotes the collection of self-adjoint operators in $V$). We then need to approximate $\hat{A}$ so as to be in $V$. However, since we eventually want to define an interval of possible values of $\hat{A}$ at $V$ we will approximate $\hat{A}$ both from above and from below. In particular, we consider the pair of operators
\ba\label{ali:order}
\delta^o(\hat{A})_V:=\bigwedge\{\hat{B}\in V_{sa}|\hat{A}\leq_s\hat{B}\}\;;\;\;\;
\delta^i(\hat{A})_V:=\bigvee\{\hat{B}\in V_{sa}|\hat{A}\geq_s\hat{B}\}
\ea
In the above equation $\delta^o(\hat{A})_V$ represents the smallest self-adjoint operator in $V$, which is spectrally larger or equal to $\hat{A}$, while $\delta^i(\hat{A})_V$ represents the biggest self-adjoint operator in $V_{sa}$, that is spectrally smaller or equal to $\hat{A}$. The process represented by $\delta^i$ is called \emph{inner dasainisation}, while $\delta^o$ represents the already encountered \emph{outer daseinisation}.

From the definition of $\delta^i(\hat{A})_V$ it follows that if $V^{'}\subseteq V$ then $\delta^i(\hat{A})_{V^{'}}\leq_s\delta^i(\hat{A})_V$. Moreover, from \ref{ali:order} it follows that:
\be
sp(\delta^i(\hat{A})_V)\subseteq sp(\hat{A}),\;\;\;\;\;sp(\delta^o(\hat{A})_V)\subseteq sp(\hat{A})
\ee
which, as mentioned above,  is precisely the reason why the spectral order was chosen.

It is interesting to note that the definition of spectral order, when applied to inner and outer daseinisation, implies the following:
\be
\delta^i(\hat{A})_{V}\leq_s\delta^o(\hat{A})_V\;\;\text{ iff }\forall r\in\Rl\;\;\;\hat{E}^{\delta^i(\hat{A})_{V}}_r\geq\hat{E}^{\delta^o(\hat{A})_{V}}_r
\ee
Since $\delta^i(\hat{A})_{V}\leq\delta^o(\hat{A})_V$, it follows that for all $r\in\Rl$
\ba
\hat{E}^{\delta^i(\hat{A})_{V}}_r:=\delta^o(\hat{E}^{\hat{A}}_r)_V\;;\;\;
\hat{E}^{\delta^o(\hat{A})_{V}}_r:=\delta^i(\hat{E}^{\hat{A}}_r)_V
\ea
The spectral family described by the second equation is right-continuous, while the first is not. To overcome this problem we define the following:
\be
\hat{E}^{\delta^i(\hat{A})_{V}}_r:=\bigwedge_{s> r}\delta^o(\hat{E}^{\hat{A}}_s)_V
\ee
Putting together all the results above we can write inner and outer daseinisation of self-adjoint operators as follows:
\ba
\delta^o(\hat{A})_V:=\int_{\Rl}\lambda d\Big(\delta^i(\hat{E}_{\lambda}^{\hat{A}})\Big)\text{ and }
\delta^i(\hat{A})_V:=\int_{\Rl}\lambda d\Big(\bigwedge_{\mu>\lambda}\delta^o(\hat{E}_{\mu}^{\hat{A}})\Big)
\ea
where the  integrals are Riemann-Stieltjes integrals\footnote{This is why right continuity was needed}.
We can now define the order-reversing and order-preserving functions as follows:
\ba
\mu_{\lambda}:\downarrow V\rightarrow \Rl\;;\;\;
V^{'}\mapsto\lambda_{|V^{'}}(\delta^i(\hat{A})_{V^{'}})=\lambda(\delta^i(\hat{A})_{V^{'}})
\ea
The order-reversing functions are defined as follows:
\ba
\nu_{\lambda}:\downarrow V\rightarrow \Rl\;;\;\;
V^{'}\mapsto\lambda_{|V^{'}}(\delta^o(\hat{A})_{V^{'}})=\lambda(\delta^o(\hat{A})_{V^{'}})
\ea

\section{Truth Values Using the Pseudo-State Object }
We are now ready to turn to the question of how truth values are assigned to
propositions which, as we have seen, are represented by daseinised
operators $\delta(\hat{P})$. For this purpose it is worth thinking
again about classical physics.  There, as previously stated,  a proposition
${A}\in\Delta$ is true for a given state $s$ iff $s\in
f_{{A}}^{-1}(\Delta)$, i.e., if $s$ belongs to those subsets
$f_{{A}}^{-1}(\Delta)$ of the state space for which the
proposition ${A}\in\Delta$ is true. Therefore, given a state
$s$, all true propositions of $s$ are represented by those
measurable subsets which contain $s$, i.e., those subsets which
have measure $1$ with respect to the measure $\delta_s$.

In the quantum case, a proposition of the form ``$A\in\Delta$'' is
represented by the presheaf $\ps{\delta(\hat{E}[A\in\Delta])}$,
where $\hat E[A\in\Delta]$ is the spectral projector for the
self-adjoint operator $\hat A$, which projects onto the subset $\Delta$ of the
spectrum of $\hat A$. On the other hand, states are represented by
the presheaves $\ps{\w}^{\ket\psi}$. As described above, within the structure of formal, typed languages,
both presheaves $\ps{\w}^{\ket\psi}$ and $\ps{\delta(\hat{P})}$
are terms of type $P\Sig$ \cite{bell}.

%
We now want to define the condition by which, for each context
$V$, the proposition $(\ps{\delta(\hat{P})})_V$ is true given
$\ps{\w}^{\ket\psi}_V$. To this end we recall that, $\ps{\w}^{\ket\psi}:= \{\ps{\w}^{\ket\psi}_V \mid{V\in\V(\Hi)}\}$
is the subobject of the spectral presheaf $\Sig$ such that at each
context $V\in\V(\Hi)$ it identifies those subsets of the Gel'fand
spectrum which correspond (through the map $\mathfrak{S}$) to the
smallest projections of that context, which have expectation value
equal to one with respect to the state $\ket\psi$, i.e., which are
true in $\ket\psi$.
On the other hand, as previously stated, the sub-presheaf $\ps{\delta(\hat{P})}$ is defined as the
subobject of $\Sig $ which, at each context $V$ is the
subset $\ps{\delta(\hat{P})}_V$ of the Gel'fand spectrum $\Sig(V)$,
which represents (through the map $\mathfrak{S}$) the projection
operator $\delta(\hat{P})_V$.
%

We are interested in defining the condition by which the
proposition represented by the subobject $\ps{\delta(\hat{P})}$ is
true given the state $\ps{\w}^{\ket\psi}$. Let us analyse this
condition for each context V. In this case, we need to define the
condition by which the projection operator $\delta(\hat{P})_V$,
associated to the proposition $\ps{\delta(\hat{P})}$ is true given
the pseudo state $\ps{\w}^{\ket\psi}$. Since at each context $V$
the pseudo-state defines the smallest projection in that context
which is true with probability one, i.e., $(\w^{\ket\psi})_V$, for
any other projection to be true given this pseudo-state, this
projection must be a coarse-graining of $(\w^{\ket\psi})_V$, i.e.,
it must be implied by $(\w^{\ket\psi})_V$. Thus, if
$(\w^{\ket\psi})_V$ is the smallest projection in $P(V)$ which is
true with probability one, then the projector $\delta^o(\hat{P})_V$
will be true if and only if $\delta^o(\hat{P})_V\geq
(\w^{\ket\psi})_V$. This condition is a consequence of the fact
that, if $\langle\psi|\hat{\alpha}\ket\psi=1$, then for all
$\hat{\beta}\geq\hat{\alpha}$ it follows that
$\langle\psi|\hat{\beta}\ket\psi=1$.

So far we have defined a `truthfulness' relation at the level of
projection operators. Through the map $\mathfrak{S}$ it is
possible to shift this relation to the level of subobjects of the
Gel'fand spectrum:
\begin{align}
\mathfrak{S}((\w^{\ket\psi})_V)&\subseteq
\mathfrak{S}(\delta^o(\hat{P})_V)\label{equ:truthvalue}\\
\ps{\w^{\ket\psi}}_V&\subseteq
\ps{\delta(\hat{P})}_V\nonumber\\
\{\lambda\in\Sig(V)|\lambda
((\delta^o(\ket\psi\langle\psi|)_V)=1\}&\subseteq
\{\lambda\in\Sig(V)|\lambda((\delta^o(\hat{P}))_V)=1\}
\end{align}
What the above equation shows is that, at the level of
subobjects of the Gel'fand spectrum, for each context $V$, a
`proposition' can be said to be (totally) true for a given pseudo-state if, and only if, the subsets of the Gel'fand
spectrum, associated to the pseudo-state are subsets of the
corresponding subsets of the Gel'fand spectrum associated to the
proposition. It is straightforward to see that if
$\delta(\hat{P})_V\geq (\w^{\ket\psi})_V$, then
$\mathfrak{S}((\w^{\ket\psi})_V)\subseteq
\mathfrak{S}(\delta(\hat{P})_V)$, since for projection operators
the map $\lambda$ takes the values 0,1 only.

We still need a further abstraction in order to work directly with
the presheaves $\ps{\w}^{\ket\psi}$ and $\ps{\delta(\hat{P})}$.
Thus we want the analogue of equation (\ref{equ:truthvalue}) at
the level of subobjects of the spectral presheaf, $\Sig$. This
relation is easily derived to be
\begin{equation}\label{equ:tpre}
\ps{\w}^{\ket\psi}\subseteq\ps{\delta(\hat{P})}
\end{equation}

Equation (\ref{equ:tpre})  shows that, whether or not a proposition
$\ps{\delta(\hat{P})}$ is `totally true' given a pseudo state
$\ps{\w}^{\ket\psi}$ is determined by whether or not the
pseudo-state is a sub-presheaf of the presheaf
$\ps{\delta(\hat{P})}$. With motivation, we can now define the
\emph{generalised truth value} of the proposition ``$A\in\Delta$''
at stage $V$, given the state $\ps{\w}^{\ket\psi}$, as:
\begin{align}\label{ali:true}
v(A\in\Delta;\ket\psi)_V&= v(\ps{\w}^{\ket\psi}
\subseteq\ps{\delta(\hat{E}[A\in\Delta])})_V             \\ \nonumber
&:=\{V^{'}\subseteq V|(\ps{\w}^{\ket\psi})_{V^{'}}\subseteq
\ps{\delta(\hat{E}[A\in\Delta]))}_{V^{'}}\}\\ \nonumber
&=\{V^{'}\subseteq
V|\langle\psi|\delta(\hat{E}[A\in\Delta])_{V^{'}}\ket\psi=1\}
\end{align}
The last equality is derived by the fact that
$(\ps{\w}^{\ket\psi})_V\subseteq \ps{\delta(\hat{P})}_V $ is a
consequence that, at the level of projection operator,
$\delta^o(\hat{P})_V\geq (\w^{\ket\psi})_V$. But, since
$(\w^{\ket\psi})_V$ is the smallest projection operator, such that
$\langle\psi|(\w^{\ket\psi})_V\ket\psi=1$, then
$\delta^o(\hat{P})_V\geq (\w^{\ket\psi})_V$ implies that
$\langle\psi|\delta^o(\hat{P})\ket\psi=1$.

The right hand side of equation (\ref{ali:true}) means that the
truth value, defined at $V$ of the proposition ``$A\in\Delta$'',
given the state $\ps{\w}^{\ket\psi}$, is given in terms of all
those sub-contexts $V^{'}\subseteq V$ for which the projection
operator $\delta(\hat{E}[A\in\Delta]))_{V^{'}}$ has expectation value
equal to one with respect to the state $\ket\psi$. In other words,
this \emph{partial} truth value is defined to be the set of all
those sub-contexts for which the proposition is totally true.

The reason why all this works is that generalised truth values defined
in this way form  a \emph{sieve} on $V$, and the set of all of
these is a Heyting algebra. Specifically:
$v(\ps{\w}^{\ket\psi}\subseteq \ps{\delta(\hat{P})})$ is a
global element defined at stage V of the subobject classifier
$\Om:=(\Om_V)_{V\in\V(\Hi)}$, where $\Om_V$ represents the set of
all sieves defined at stage V.
The set of global elements of  $\Omega$ is denoted $\Gamma(\Omega)$ and it forms a Heyting algebra.
%
\section{Example}
We will consider again the four-dimensional Hilbert space $\Cl^4$. We now want to define the truth values of the proposition  $S_z\in[-3, -1]$ which has corresponding projector operator $\hat{P}:=\hat{E}[S_z\in[-3, -1]]=|\phi\rangle\langle\phi|$. This proposition is equivalent to the projection operator $\hat{P}_4$, defined in previous examples.\\
The state we will consider will be $\psi=(1,0,0,0)$ with respective operator $|\psi\rangle\langle\psi|=\hat{P}_1$. First of all we will consider the context $V:=lin_{\Cl}(\hat{P}_1, \hat{P}_2, \hat{P}_3,\hat{P}_4)$. Since $\hat{P}_4\in V$ it follows that $\delta^o(\hat{P}_4)_V=\hat{P}_4$, but $\langle\psi|\hat{P}_4|\psi\rangle=0$, thus we need to go to smaller contexts. In particular, only the contexts in which $\delta^o(\hat{P}_4)$ is an operator which is implied by $\hat{P}_1$ will contribute to the truth value. We thus obtain
\ba
v(\ps{\w}^{\ket\psi}\subseteq\ps{\delta(\hat{P}_4)})_V:=\{V^{'}\subseteq
V|\langle\psi|\delta(\hat{E}[A\in\Delta])\ket\psi=1\}
=\{V_{\hat{P}_{i\in\{2,3\}}}, V_{\hat{P}_{2}\hat{P}_{3}}\}
\ea
On the other hand for sub contexts we obtain
\ba
v(\ps{\w}^{\ket\psi}\subseteq\ps{\delta(\hat{P}_4)})_{V_{\hat{P}_1}}:=\{\emptyset\}\;\;\;&,&\;\;\;
v(\ps{\w}^{\ket\psi}\subseteq\ps{\delta(\hat{P}_4)})_{V_{\hat{P}_2}}:=\{V_{\hat{P}_2}\}\\
v(\ps{\w}^{\ket\psi}\subseteq\ps{\delta(\hat{P}_4)})_{V_{\hat{P}_3}}:=\{V_{\hat{P}_3}\}\;\;\;&,&\;\;\;
v(\ps{\w}^{\ket\psi}\subseteq\ps{\delta(\hat{P}_4)})_{V_{\hat{P}_4}}:=\{\emptyset\}\nonumber\\
v(\ps{\w}^{\ket\psi}\subseteq\ps{\delta(\hat{P}_4)})_{V_{\hat{P}_1, \hat{P}_2}}:=\{V_{\hat{P}_2}\}\;\;\;&,&\;\;\;
v(\ps{\w}^{\ket\psi}\subseteq\ps{\delta(\hat{P}_4)})_{V_{\hat{P}_1, \hat{P}_3}}:=\{V_{\hat{P}_3}\}\nonumber\\
v(\ps{\w}^{\ket\psi}\subseteq\ps{\delta(\hat{P}_4)})_{V_{\hat{P}_1, \hat{P}_4}}:=\{\emptyset\}\;\;\;&,&\;\;\;
v(\ps{\w}^{\ket\psi}\subseteq\ps{\delta(\hat{P}_4)})_{V_{\hat{P}_2, \hat{P}_3}}:=\{V_{\hat{P}_2, \hat{P}_3},V_{\hat{P}_2},V_{\hat{P}_3}\}\nonumber\\
v(\ps{\w}^{\ket\psi}\subseteq\ps{\delta(\hat{P}_4)})_{V_{\hat{P}_2, \hat{P}_4}}:=\{V_{\hat{P}_2}\}\;\;\;&,&\;\;\;
v(\ps{\w}^{\ket\psi}\subseteq\ps{\delta(\hat{P}_4)})_{V_{\hat{P}_3, \hat{P}_4}}:=\{V_{\hat{P}_3}\}\nonumber
\ea
The collection of all these local truth values represents the global truth value $v(\ps{\w}^{\ket\psi}\subseteq\ps{\delta(\hat{P}_4)})$.

\section{Conclusion and Outlook}
In this review paper we have outlined, with examples, the topos formulation of quantum theory,  which suggests a more realist interpretation of the theory. Such an interpretation is preferable since it overcomes the conceptual difficulties related to the notion of closed system and the Kochen-Specker no-go theorem inherent in the standard Copenhagen interpretation of quantum theory.

However, a radical new way of thinking about what a theory of physics \emph{is}, emerges. Consequently, a different interpretation of the concepts of space, time and matter is required.

A connection with a possible theory of quantum gravity was done in \cite{46ah}  where a formulation of history quantum theory was carried out in terms of topos theory. This reformulation is very important, since it allows the possibilities of defining any quantum statements about four-metrics.

In particular, in this new topos approach of history theory it has been shown that Heyting-algebra valued truth values can be assigned to any history proposition, i.e., it is no longer necessary to consider just `consistent' sets of propositions. This is an advantage over the older consistent-history formalism in which the process of choosing which consistent set of history propositions to employ when defining quantum statements, is really problematic.

Therefore, the topos formulation of history theory sets the stage for a framework in which truth values can be assigned to \emph{any} proposition about spacetime.

Recently in \cite{probabilities} it was shown how probabilities can be described in terms of truth values in a given
topos. In such a formulation logical concepts are seen as fundamental, while probabilities become derived
concepts. This approach to probability theory allows for a new type of non-instrumentalist interpretation that might be particularly appropriate in those schemes which interpret probabilities as propensities.

Both the topos version of quantum theory, history theory and probabilities, are only the first steps towards a theory of quantum gravity in terms of topos theory. A lot of work is still needed. However, the prescription of how a theory of quantum gravity should be derived, is the same as the one used for reformulating quantum theory and history quantum theory in the language of topos theory. In particular, these theories are the result of an interplay between four main ingredients:
\begin{enumerate}
\item[1.] The physical system under consideration.
\item[2.] The type of theory one sets out to analyse (classical or quantum).
\item[3.] The corresponding correct topos in which to express the theory. The choice of such a topos will depend on the theory type and on the system under consideration.
\item[4.] The formal language or underlying logic associated to the system.
\end{enumerate}
A theory of physics is then identified with finding a representation, in a certain topos, of the formal language that is attached to the system.
This strategy revealed itself successful, for both quantum theory and history theory, with advantages and enrichment over the standard formulations of the theories in both cases.  The hope is that this same strategy can reveal itself fruitful for defining a possible quantum theory of gravity.

\bigskip
\textbf{Acknowledgements. }I would like to thank C.~Isham, A.~D\"oering for valuable discussions and support. This work was supported by the Perimeter Institute of Theoretical Physics Waterloo Ontario.

\end{document}